\documentclass[aps,prl,twocolumn,a4paper,superscriptaddress]{revtex4}

\usepackage[paperwidth=215mm,paperheight=300mm,centering,hmargin=1.6cm,vmargin=2cm]{geometry}
\usepackage{multibbl}
\usepackage{graphicx,amsmath,SIunits}
\usepackage{latexsym,stmaryrd}
\usepackage[sort&compress]{natbib}
\usepackage{color}
\bibpunct{[}{]}{,}{n}{}{,}

\makeatletter
\newcommand*{\citenst}[2][]{%
  \begingroup
  \let\NAT@mbox=\mbox
  \let\@cite\NAT@citenum
  \let\NAT@space\NAT@spacechar
  \let\NAT@super@kern\relax
  \renewcommand\NAT@open{[}%
  \renewcommand\NAT@close{]}%
  \citep{#2}%
  \endgroup
}
\makeatother

\renewcommand{\figurename}{\textbf{Figure}}

\begin{document}

\preprint{APS/123-QED}

\title{Injection locking in an optomechanical coherent phonon source}

\author{G. Arregui}
\affiliation{Catalan Institute of Nanoscience and Nanotechnology (ICN2), CSIC and The Barcelona Institute of Science and Technology, Campus UAB, Bellaterra, 08193 Barcelona, Spain}
\affiliation{Dept. de F\'{i}sica, Universitat Autonoma de Barcelona, 08193 Bellaterra, Spain}
\author{M. F. Colombano }
\affiliation{Catalan Institute of Nanoscience and Nanotechnology (ICN2), CSIC and The Barcelona Institute of Science and Technology, Campus UAB, Bellaterra, 08193 Barcelona, Spain}
\affiliation{MIND-IN2UB, Departament d'Enginyeria Electr\`onica i Biom\`edica, Facultat de F\'isica, Universitat de Barcelona, Mart\'i i Franqu\`es 1, 08028 Barcelona, Spain}
\author{J. Maire}
\affiliation{Catalan Institute of Nanoscience and Nanotechnology (ICN2), CSIC and The Barcelona Institute of Science and Technology, Campus UAB, Bellaterra, 08193 Barcelona, Spain}
\author{A. Pitanti}
\affiliation{NEST Lab., CNR - Istituto di Nanoscienze and Scuola Normale Superiore, Piazza San Silvestro 12, 56217 Pisa, Italy}
\author{N. E. Capuj}
\affiliation{Depto. F\'{i}sica, Universidad de La Laguna, 38200 San Crist\'{o}bal de La Laguna, Spain}
\affiliation{Instituto Universitario de Materiales y Nanotecnolog\'{i}a, Universidad de La Laguna, 38071 Santa Cruz de Tenerife, Spain}
\author{A. Griol}
\affiliation{Nanophotonics Technology Center, Universitat Polit\`ecnica de Valencia, 46022 Valencia, Spain}
\author{A. Mart\'{i}nez}
\affiliation{Nanophotonics Technology Center, Universitat Polit\`ecnica de Valencia, 46022 Valencia, Spain}
\author{C. M. Sotomayor-Torres}
\affiliation{Catalan Institute of Nanoscience and Nanotechnology (ICN2), CSIC and The Barcelona Institute of Science and Technology, Campus UAB, Bellaterra, 08193 Barcelona, Spain}
\affiliation{ICREA - Instituci\'o Catalana de Recerca i Estudis Avan\c{c}ats, 08010 Barcelona, Spain}
\author{D. Navarro-Urrios}
\email{dnavarro@ub.edu}
\affiliation{MIND-IN2UB, Departament d'Enginyeria Electr\`onica i Biom\`edica, Facultat de F\'isica, Universitat de Barcelona, Mart\'i i Franqu\`es 1, 08028 Barcelona, Spain}

\date{\today}

\small

\begin{abstract}
Spontaneous locking of the phase of a coherent phonon source to an external reference is demonstrated in an optomechanical oscillator based on a self-triggered free-carrier/temperature limit cycle. Synchronization is observed when the pump laser driving the mechanical oscillator to a self-sustained state is modulated by a radiofrequency tone. We employ a pump-probe phonon detection scheme based on an independent optical cavity to observe only the mechanical oscillator dynamics. The lock range of the oscillation frequency, i.e., the Arnold tongue, is experimentally determined over a range of external reference strengths, evidencing the possibility to tune the oscillator frequency for a range up to 350 kHz. The stability of the coherent phonon source is evaluated via its phase noise, with a maximum achieved suppression of 44 dBc/Hz at 1kHz offset for a 100 MHz mechanical resonator. Introducing a weak modulation in the excitation laser reveals as a further knob to trigger, control and stabilise the dynamical solutions of self-pulsing based optomechanical oscillators, thus enhancing their potential as acoustic wave sources in a single layer silicon platform.
\end{abstract}

 \pacs{(42.70.Qs, 03.65.Vf, 42.25.Dd, 42.25.Fx, 46.65.+g,  42.25.Bs, 78.67.−n)}

\maketitle

Enabled by progress in nanofabrication, the enhanced coupling of optical and mechanical degrees of freedom in purposely-engineered micro/nanoscale structures has provided physicists with the possibility to observe new physical phenomena in both the classical and quantum regime~\cite{cavityoptomechanics}. In particular, their interaction in a mechanically compliant and laser-driven optical cavity has led to the design of cavity-optomechanical systems that enables both cooling~\cite{chan} and amplification of mechanical motion~\cite{amplification} by mere selection of the driving laser wavelength. This versatility can be used for tasks spanning the exploration of the quantum nature of mesoscopic objects to on-chip signal processing. Among the applications are a new category of self-sustained radio-frequency oscillators called optomechanical oscillators (OMOs)~\cite{vahalaoscillator} which, by construction, also serve as self-sustained coherent phonon sources and may be used as such. Since any physical property in a solid-state system depends on the exact positioning of the atoms, coherent acoustic phonons are extremely suited to dynamically manipulate electric~\cite{electriccontrol1,electriccontrol2}, optical~\cite{opticcontrol1,opticcontrol2} and magnetic~\cite{magneticcontrol1,magneticcontrol2} properties of matter. Acoustic wave sources are therefore an enabling technology across physics  as well as the first building block for information processing with phonons~\cite{painterwaveguide,phononsinfo}\\

Even though several strategies to induce mechanical lasing in optomechanical devices have been proposed~\cite{phononlasing2,phononlasing3}, OMOs based on a single optical and mechanical mode are probably the only practical small-footprint implementation~\cite{ghorbel,laurita}. They can trigger an oscillation from a direct continuous source without needing feedback electronics by using mechanisms such as the retarded radiation-pressure force~\cite{phononlasing2}, the back-action induced by the bolometric light force~\cite{bolometric} or by coupling the optomechanical system to carrier/thermal self-sustained cyclic dynamics~\cite{scireports}. In these cases, their all-optical operation, ease of miniaturization, low power consumption and scalability make these oscillators potential candidates as optically-driven phonon sources~\cite{daniphononsources} and a possible replacement to conventional quartz-based oscillators in specific RF-photonic communication and sensing applications such as optical down-conversion~\cite{downconversion} or mass sensing~\cite{OMOsensing}. Although the reported output stability of OMOs approaches state-of-the-art optoelectronic oscillators~\cite{chiweiwong}, their performance is often degraded by mechanical effects such as slow frequency drift~\cite{drift}, intrinsic~\cite{freqstab} or thermomechanical noise~\cite{thermomechanical} and by instabilities occurring at large displacement amplitudes~\cite{largeamplitude}. Additional post-fabrication tunability of their phase and frequency as well as their noise characteristics are therefore critical for their improved performance as both acoustic wave sources and carriers in microwave photonics. \\

\begin{figure}
\centering\includegraphics[width=\columnwidth]{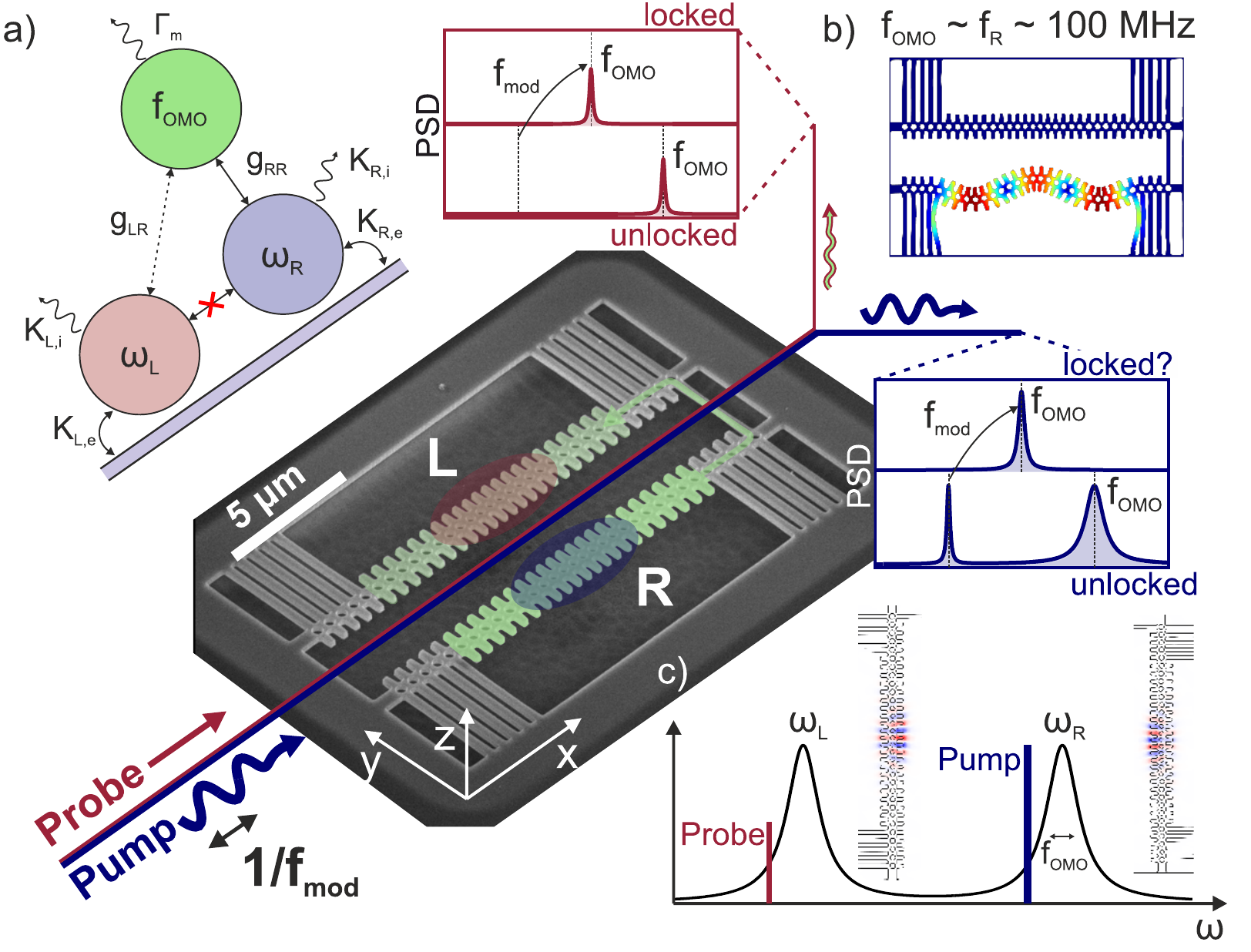}
    \caption{ \label{2} \textbf{Optomechanical system and detection scheme.} (a) Illustration of the optomechanical setting explored. Two optical cavities of frequencies $\omega_R$ (blue) and $\omega_L$ (red), with $\omega_R-\omega_L\gg\kappa_{L,R}$, can be driven via a common bus waveguide. A single mechanical mode of interest (green) is coupled optomechanically to both optical cavities, although the coupling strength is highly asymmetric, i.e., $g_{RR}\gg g_{LR}$. Cavity R is optically driven, leading to optomechanical self-oscillation of the mechanical mode at frequency $f_{OMO}$. This system is realized with two close-by silicon optomechanical crystal nanobeams that are connected by a mechanical link. Both beams support optical cavitiy modes that are spectrally and spatially isolated (c), as shown in the central SEM image. A flexural mode of the right beam (b) is selected by pumping (blue path) the right optical cavity mode at a wavelength and power where phonon lasing at frequency $f_{OMO}\sim f_m \sim 100 MHz$ occurs. Due to the link, the mechanical mode also couples weakly to the left optical cavity, which is adressed via a probe optical signal (red path). The optomechanical transduction of the right (left) cavity exhibits a strong (weak) peak at the oscillation frequency $f_{OMO}$. If the pump laser is intensity-modulated at frequency $f_{mod}$, the PSD of transmittted pump (probe) light has (does not have) an additional peak at $f_{mod}$. Tuning $f_{mod}$ can lead to injection locking. While the presence of a single peak is a necessary condition for injection-locking, it is not a sufficient one in the pump PSD (blue box). The presence of a peak at $f_{mod}$ in the PSD of the probe laser is however an unambiguous evidence of injection locking (red box).}
\label{fig:1}
\end{figure}

The behaviour of a regenerative oscillator driven by an extra external oscillatory signal with a frequency close the the oscillationl frequency has been explored in a variety of systems such as lasers~\cite{laserlock1,laserlock2} or masers~\cite{micromaser}, electrical tank circuits~\cite{electroniclock1,electroniclock2}, organ pipes~\cite{organpipes}, nonlinear mechanical resonators~\cite{weig,micromechanical}, superconducting circuits~\cite{superconducting} and human circadian rythms~\cite{circadian}. These apparently dissimilar systems all share their tendency towards an adjustment of the oscillator's rythm to the one externally imposed, as originally proposed by Adler~\cite{adler}. In recent years, the intrinsic nonlinearity of the optomechanical interaction has been harnessed to explore the spontaneous locking of an oscillator to a reference signal delivered via the driving optical field~\cite{opticaldrive1,opticaldrive2,opticaldrive3,chiweiwong}, via electrical capacitive actuation~\cite{pitanti, optica} or by mechanical actuation with propagating acoustic waves~\cite{ILacoustic}. In all these cases the system is driven to a mechanical lasing state by dynamical back-action, which requires a large optomechanical cooperativity~\cite{cavityoptomechanics}. In this article, we demonstrate spontaneous frequency locking of the coherent mechanical motion of a one-dimensional silicon optomechanical crystal (OMC) brought to a self-sustained state using the anharmonic modulation of the radiation pressure force due to the activation of a self-pulsing (SP) mechanism~\cite{SP}. We use an independent optical cavity the radiofrequency spectrum of which is solely affected by the self-sustained mechanical dynamics. In this way, we overcome a drawback of the transduction schemes in previous experiments, which often prevented an unambiguous assesment of the instantaneous amplitude of the mechanical oscillator due to the direct measurement of the reference oscillatory signal. \\

The particular system investigated here is a pair of close-by one-dimensional OMCs fabricated in a silicon-on-insulator wafer, as shown in the central SEM micrograph of Fig.~\ref{fig:1}. We investigate the fundamental optical cavity mode of each OMC at wavelengths of 1531 nm (right-R) and 1550 nm (left-L), respectively, whose eigenfields are shown in Fig.~\ref{fig:1}(c). Fabrication disorder distorts the nominal y-symmetry of the single OMC optical field profiles and naturally provides the spectral detuning between the two optical cavities, preventing any optical crosstalk as shown in the schematic of Fig.~\ref{fig:1}. In addition, the same distortion leads to a high vacuum optomechanical coupling rate $g_{o}$ between the optical resonances and the mechanical flexural modes with three antinodes along the x direction, that would otherwise be optomechanically dark. For the right OMC, this mechanical mode [Fig.~\ref{fig:1}(b)] displays a frequency of $f_R=100.37$ MHz and a calculated value of $g_{o,RR}$ = 2$\pi\cdot$514 kHz. The precise geometry and reasoning behind the clamping and interconnect structure has been described elsewhere~\cite{synchro}, but it essentially allows the simultaneous optical excitation of the two optical modes as well as controlled phonon leakage from one beam to its neighbour. The mechanical eigenstructure of identical nanobeams would exhibit the symmetry of the system and would induce a perfect hybridization of the two mechanical modes into a symmetric and antisymmetric coupled modes. The mentioned fabrication disorder breaks the as-designed symmetry and the original string-like modes weakly hybridize as a result. However the two-mode picture with in-phase and anti-phase oscillation still applies (see Supplementary Information). As a result of this weak coupling, a small part of the energy of the mechanical mode supported by the right beam (imperceptible in Fig.~\ref{fig:1}(b)) resides in the left beam, which leads to a small calculated cross-coupling term of $g_{o,LR}$ = 2$\pi\cdot$8 kHz between the mechanical mode of the right OMC and the optical cavity of the left OMC, two orders of magnitude smaller than the direct term $g_{o,RR}$. The mechanical dynamics of the right OMC are controlled via a strong pump laser (dark blue in Fig.~\ref{fig:1}) that couples light into the right optical cavity, eventually driving the mechanical mode into a phonon lasing state at frequency $f_{OMO}$. In this coherent state, the mechanical dynamics are accessed by inspecting the optomechanical transduction of a weak probe optical signal (dark red in Fig.~\ref{fig:1}) addressed on the left cavity, a signal that originates from the cross coupling term $g_{o,LR}$. When the pump laser is modulated with a weak RF tone at frequency $f_{mod}$ far from the oscillation frequency $f_{OMO}$, this imprints an additional peak in the power spectral density (PSD) of the pump laser, but does not change the PSD of the probe laser, which, by construction, only shows the peak at $f_{OMO}$. Bringing $f_{mod}$ sufficiently close to $f_{OMO}$ can lock the OMO to $f_{mod}$, which would lead to a single peak. However, the direct spectral observation of the modulation imposed onto the pump when reading-out the pump after interaction with the sample complicates the unambiguous attribution of the presence of a single peak to injection locking at a particular modulation strength and frequency, as represented via the blue box in Fig.~\ref{fig:1}. To discard the possibility that the external modulation is strong enough to quench the initially prepared lasing dynamical state, a direct determination of the mechanical dynamics is required. To this end the probe cavity is used. When the probe laser is analysed at the output, the presence of a peak at $f_{mod}=f_{OMO}$ of similar magnitude to that observed at $f_{OMO}$ in the unlocked case is an unambiguous spectral evidence of injection locking, as depicted in the red box of Fig.~\ref{fig:1}.\\

\begin{figure*}
\centering\includegraphics[width=\textwidth]{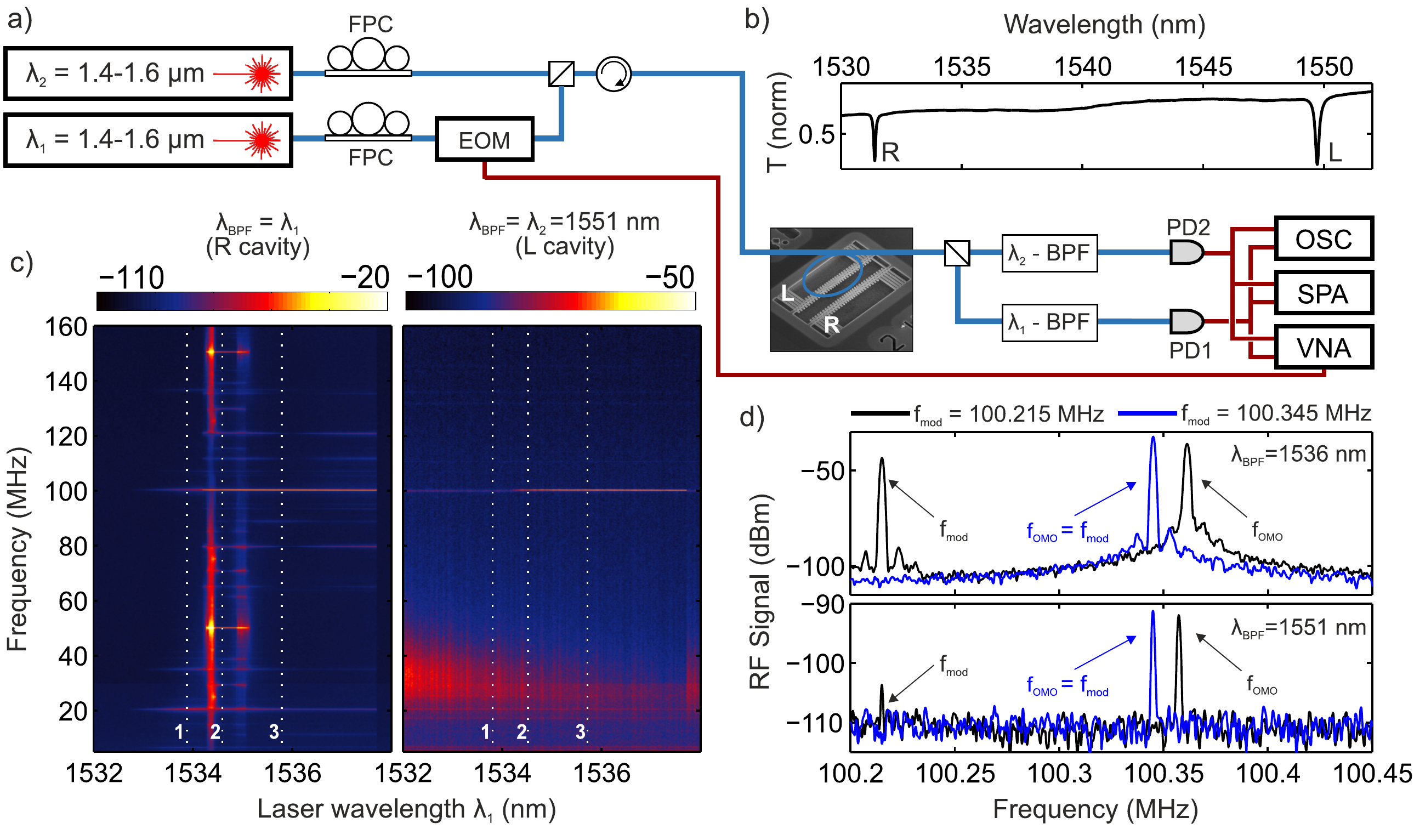}
    \caption{ \label{2} \textbf{Experimental setup and device characterization.} (a) The main optical signals are derived from two external-cavity diode laser and sent into a tapered  microloop optical fibre to evanescently couple light into the silicon optomechanical device. The optical signal from the OMC is collected by the same fibre taper and equally split in two, either in transmission or in reflection using an in-fiber circulator. The optical signals in both arms are band-pass filtered by narrow wavelength filters ($\lambda$-BPF) and impinge on two fast photodetectors (PD). The electrical signals are sent to a spectrum analyzer (SPA), a vector network analyzer (VNA) or an oscilloscope (OSC) for characterization. EOM stands for electro-optic modulator and  FPC for fibre polarization controller. (b) Optical transmission spectrum, exhibiting two optical modes at $\lambda_R$=1531 nm and $\lambda_L$=1550 nm. (c) Colormaps of the power spectral density (PSD) of the transmitted light when sweeping the laser (1) wavelength driving the optical resonance R at high power while another laser (2) weakly probes the resonance L; (left) $\lambda_{BPF}$ = $\lambda_1$ and (right) $\lambda_{BPF}$ = $\lambda_2$. The dashed white lines highlight (1) the thermal transduction, (2) $M=2$ mechanical lasing and (3) $M=1$ mechanical lasing states. (e) Within the configuration achieved in dashed line 3, adding a weak intensity-modulation via the EOM results in a peak in the PSD when its frequency $f_{mod}$ is far from the phonon lasing frequency $f_{OMO}$ (black signal), but locks the oscillator frequency for small detunings (blue signal). When the adressed signal is the probe, the spectra reproduce the ones schematically given in Fig.~\ref{fig:1}.}
\label{fig:2}
\end{figure*}

Measurements are performed at ambient conditions by bringing a tapered microloop optical fibre between the two structures, coupling light propagating in the fiber to the fundamental optical resonance of both beams [Fig.~\ref{fig:2}(b)]. The measured optical resonances have quality factors of $Q_{R}=1.6\cdot 10^4$ and $Q_{L}=8.1\cdot 10^3$, respectively, which are much lower than their calculated values $Q_{calc}=6\cdot 10^5$ due to the presence of surface roughness. Simultaneous coupling to both cavities is achieved via two tuneable near-infrared lasers (Yenista TUNICS T100S-HP) and an in-fiber combiner. The transmitted/reflected power is split and a Fabry-P\'erot bandpass filter (BPF) in each arm is used to measure independently the light coming from the chosen laser, allowing the dynamics associated to both optical resonances to be read-out separately. The signals are fed to a spectrum analyzer (SPA), to a 5 GHz oscilloscope and/or to a vector network analyzer (VNA). The latter provides an AC voltage $V_{AC}$ to an electro-optic modulator (EOM) and modulates the intensity of the input laser light, as well as measuring the magnitude and phase of the scattering response ($S_{21}$) of the system at the excitation frequency. The EOM is set at the quadrature point $V_{DC}=0.5V_{\pi}$ to suppress higher harmonics, with the half-wave voltage given by $V_{\pi}=6.7$ V. The full measurement setup is shown in Fig.~\ref{fig:2}(a). \\

Sweeping the laser (1) wavelength at high CW power from the blue-detuned side of the resonance at $\lambda_R$= 1531 nm to $\lambda$=1538 nm leads to multiple dynamical states of the right OMC. For large detunings and relatively low intracavity photon numbers $n_c$, the incoherent mechanical motion activated by the thermal Langevin force imprints a modulation around the transmission DC value. The measured PSD is highlighted with dashed line 1 in the left colormap of Fig.~\ref{fig:2}(c) and evidences the presence of several mechanical modes. In particular, the motion of the aforementioned mechanical mode at $f_{R}=100.37$ MHz is transduced, with a measured natural linewidth of $\Gamma_m=0.15$ MHz, i.e. $Q_m\sim 670$. As the wavelength, and therefore $n_c$, increases, the absorption in the tightly-confined optical mode volume leads to the generation of free-carriers and to heating of the lattice. Since the refractive index of silicon depends on both temperature and free-carrier density, the cavity dynamics are governed by a system of non-linear coupled differential equations. The dynamical interplay between free-carrier dispersion (FCD) and the thermo-optic (TO) effect leads to an anharmonic oscillation of the cavity resonance around the laser line at a frequency $\nu_{SP}$, which is itself a function of the driving parameters. As a consequence, further sweeping the wavelength tunes $\nu_{SP}$ and whenever any of the harmonics of the SP-induced modulation of $n_c$ approaches the mechanical frequency $f_m$, the mechanical resonator is pumped via the radiation-pressure force. Then the mechanical oscillation generally entrains the SP, forcing $\nu_{SP}$ to be a simple fraction $M$ of $f_m$~\cite{naturecomm}. This can be observed for $M$=2  and $M$=1 in the white dashed lines 2 and 3 of Fig.~\ref{fig:2}(c). Detailed information on the SP and the induced phonon lasing is given in the Supplementary Information and elsewhere~\cite{scireports,naturecomm}. The interest of using the neigbouring OMC can be foreseen on the right colorplot of Fig.~\ref{fig:2}(c), where the spectrum of laser (2) probing cavity L is shown as the wavelength of laser (1) is swept. The data shown in both panels is acquired simultaneously. They evidence that the amount of leaked energy coming from the oscillation of the right nanobeam and the resulting cross coupling term g$_{o,LR}$ enables reading the mechanical dynamics whilst filtering out other phenomena. Indeed, such a scheme disentangles the read-out of the mechanics from the cyclic dynamics of the free-carriers and temperature, the spatially diffusive nature and characteristic length-scales of which minimize their contribution to the neighbour cavity. It is observed that the modulation peak at 100.37 MHz is present for both an M1 and M2 phonon lasing states -dashed lines 2 and 3 in Fig.~\ref{fig:2}(c)- without the presence of previous harmonics associated to the SP dynamics. The readout laser (2) power is set to a value sufficiently high to transduce the incoming mechanical signal, but below its own SP Hopf bifurcation~\cite{hopf}. The broad spectral feature spanning 20-40 MHz across the right colormap of Fig.~\ref{fig:2}(c) is a fingerprint of proximity to such bifurcation and is used as a sign of stability in the optical coupling conditions throughout the measurements. We therefore experimentally satisfy the detection scheme depicted in Fig.~\ref{fig:1}, which opens the door to act upon the dynamical state of the complex self-pulsing/lasing state of the right cavity and observing the effect on the mechanical resonator dynamics via the left cavity. This experimental scheme was previously employed in~\cite{synchro}, where both mechanical modes were brought to lasing and their motion synchronized. Further details are found in the Supplementary Information.\\

\begin{figure}
  \centering\includegraphics[width=\columnwidth]{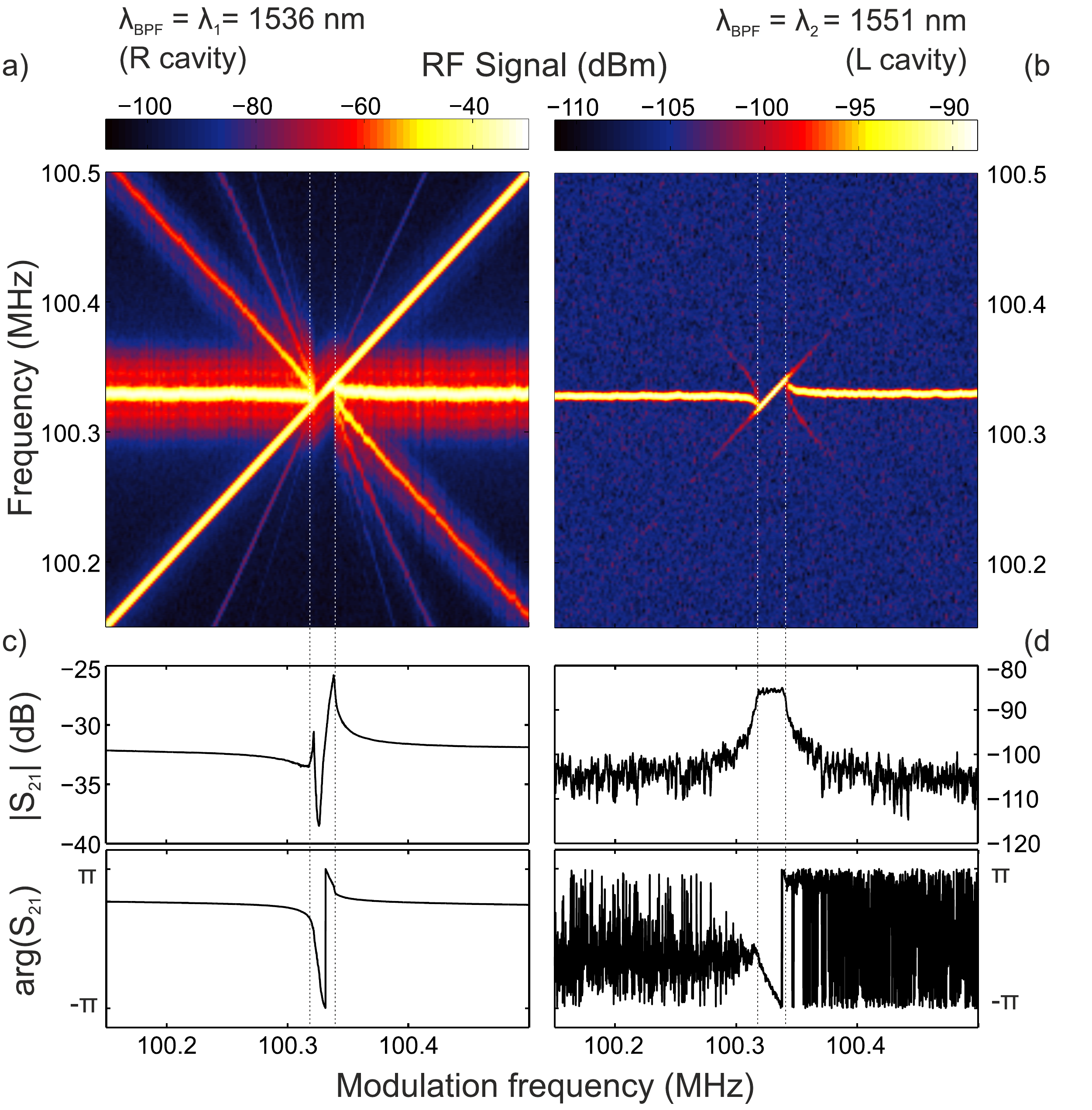}
    \caption{ \textbf{Spectral-domain evidence of injection locking to an external modulation.} (Color online) (a) Color plot of the power spectral density (PSD) of the transmitted light of the laser driving the right beam fundamental optical resonance at $\lambda_1$ = 1536 nm with $P_{in}$ = 3 mW when changing the modulation frequency in steps of 100 Hz. In a given range the phonon lasing peak synchronizes to the modulation tone.  (b) Equivalent measurement by observing light from the second laser that weakly probes the optical mode at $\lambda_2$ = 1550 nm of the left beam, allowing only the observation of the mechanical dynamics of the right beam. The coherent complex scattering response of the system measuring $S_{21}$ in the VNA is depicted for the filter aligned with the laser driving the right (c) and the left (d) optical modes.}
\label{fig:3}
\end{figure}

We focus now on the particular situation represented by dashed line 3 in Fig.~\ref{fig:2}(c), where the right beam is pumped by laser (1) to the $M=1$ mechanical lasing state at $f_{OMO,0}=100.36$ MHz and we modulate the power of the laser at a frequency $f_{mod}$ slightly detuned from the mechanical frequency ($V_{AC}=0.004V_{\pi}$, $f_{mod}=100.215$ MHz). Note that the free-running self-sustained oscillation frequency $f_{OMO,0}$ in the absence of external  modulation is usually smaller than the natural frequency $f_{R}$ due to thermomechanical softening~\cite{modetuning}. In the transmitted signal RF spectrum two distinct peaks at the oscillator frequency $f_{OMO}\sim f_{OMO,0}$ and at the modulation frequency $f_{mod}$ are observed, as shown by the top black curve of Fig.~\ref{fig:2}(d). For a smaller frequency difference, i.e., $f_{OMO,0}-f_{mod}=0.015$ MHz, only one peak at $f_{mod}$ is observed (blue curve), which is a first fingerprint of injection locking. However this is unambiguously confirmed by employing the phonon detection scheme previously described. With the laser (2) as the probe set to be in resonance with the optical mode of the left beam and perfectly transmitted by the BPF, we are able to observe the mechanical lasing state of the neighbour cavity. When the modulation to laser (1) is on (bottom of Fig.~\ref{fig:2}(d)), a peak at the modulation frequency $f_{mod}$ becomes prominent only when the mechanical lasing is injection-locked to the reference (blue curve) and $f_{OMO}=f_{mod}$. For a larger frequency difference (black curve), only the transduction of the mechanical lasing at $f_{OMO}$ is strong. In this modulation conditions, the presence of the small peak at the modulation frequency evidences the onset of non-trivial mechanical dynamics of the lasing cavity. Indeed, a possible contribution from free-carrier/temperature dynamics at $f_{mod}$ has already been discarded by the fact that only the mechanical peak is detected in that configuration when the lasing cavity is in $M=2$ [Fig.~\ref{fig:2}(c)].\\

\begin{figure}
 \centering \includegraphics[width=\columnwidth]{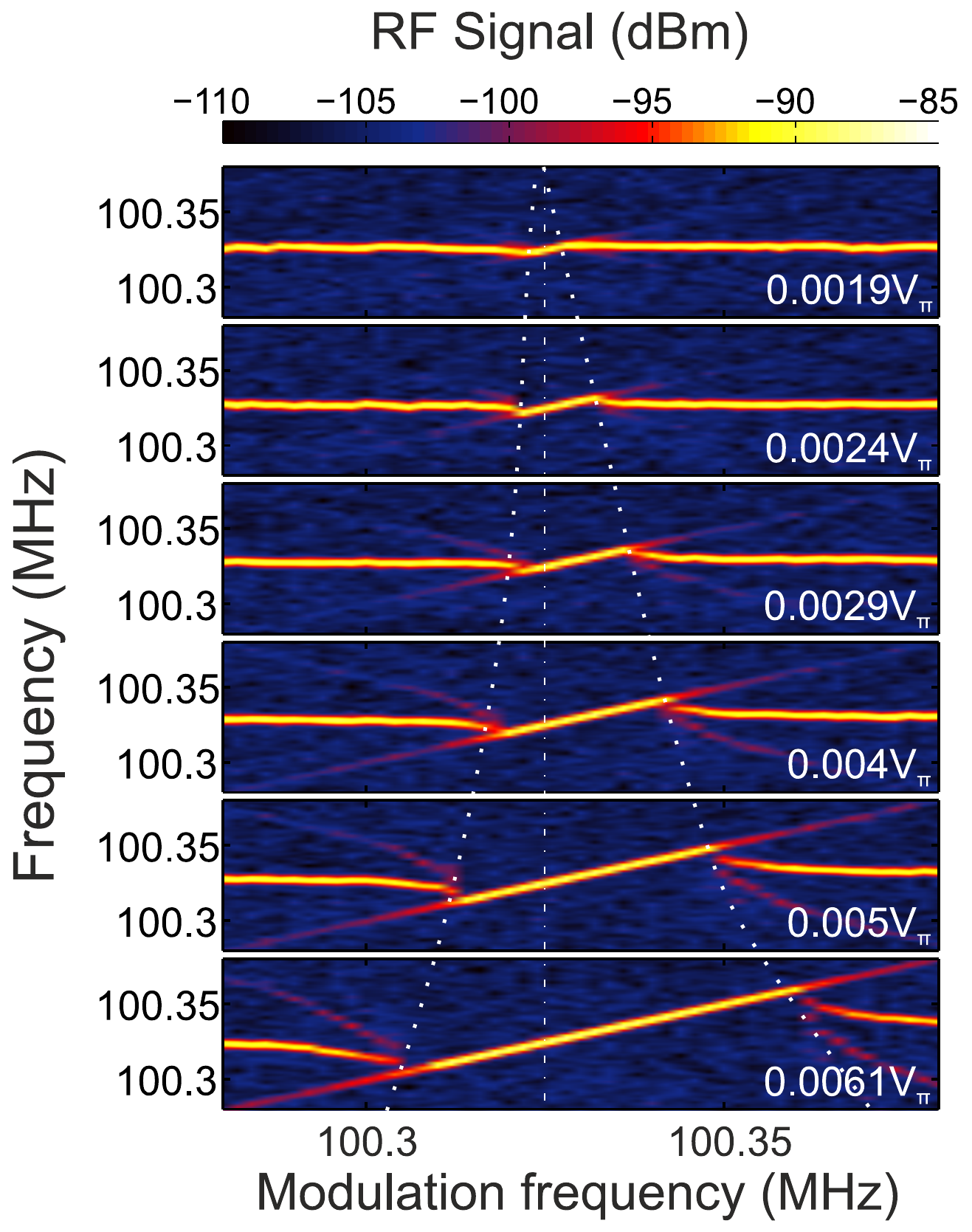}
    \caption{ \textbf{Phonon lasing oscillation frequency.} (Color online) Colormaps of the power spectral density (PSD) of light transmitted at the wavelength of the probe laser for modulation amplitudes from $V_{AC}$ = 0.0019$V_{\pi}$ to $V_{AC}$ = 0.0061$V_{\pi}$. The dashed white lines are guides to the eye and help to visualise the asymmetric growth of the Arnold tongue with respect to the unlocked oscillation frequency at $f_{OMO,0}$=100.322 MHz.}
\label{fig:4}
\end{figure}

The range of modulation frequencies for which injection locking occurs, i.e., the Arnold tongue~\cite{arnold}, is investigated by performing measurements such as those shown in Fig.~\ref{fig:2}(d) while step-changing the excitation frequency from lower to higher frequencies. The results for $V_{AC}=0.004V_{\pi}$, correspoding to a VNA tone of -5 dBm, are depicted as a colormap in Fig.~\ref{fig:3}(a-b), with the BPFs centered in $\lambda=1536$ nm and $\lambda = 1551$ nm, respectively. Note that during the acquisition of Fig.~\ref{fig:3}(a-b), $f_{OMO}$=100.322 MHz in the absence of modulation is slightly lower than previously shown, which is due to a change in the optical coupling condition. Panel (a) clearly evidences the presence of an injection-locked region spanning 23 kHz, confirmed by the measurement performed on the other cavity (panel (b)). The absence of the directly transmitted and modulated light allows the evolution of the dynamics of the mechanical oscillator to be observed. The first remarkable observation is that prior to injection-locking, the mechanical response exhibits a peak at the modulation frequency $f_{mod}$, indicating that the trajectory of the oscillator becomes non sinusoidal, as was already pointed out in Fig.~\ref{fig:2}(d). Moreover, approaching the edges of the lock range, the peak at the mechanical lasing is dragged towards $f_{mod}$, i.e., frequency pulling occurs, which is a typical feature of self-sustained oscillators driven by an external source~\cite{adler,pikovsky}. In addition to aquiring the PSD signal of the transmitted light, the complex scattering parameter $S_{21}$ of the system is measured by the VNA. Fig.~\ref{fig:3}(c-d) show its magnitude and phase when measured with the filter pass-band in resonance with either laser. The signal obtained by measuring at the pump laser wavelength is complex since both the modulation from the directly transmitted light and the modulation from the effect of the frequency tone on the cavity dynamics can interfere. However, the amplitude response observed in Fig.~\ref{fig:3}(d) is simpler and shows that the oscillation amplitude of the mechanical oscillator remains constant throughout the lock range, at least for the corresponding modulation amplitude, and that the RF tone appearing at the borders outside that range follows the characteristic lineshape of a driven harmonic oscillator. Furthermore, the phase response $\Delta\phi$ is characterized by an evolution of approximately $\pi/2$ across the lock range, which is another standard feature of injection-locked electric tank oscillators~\cite{electroniclock1}. The PSD signal acquired for several modulation amplitudes $V_{AC}$, now only measuring with the filter transmitting the laser light outcoupled from the probe cavity, is depicted in Figure~\ref{fig:4}. This set of measurements clearly demonstrates the increasing size of the Arnold tongue with modulation amplitude and its asymmetric growth. This asymmetry, which is not typical in other injection-locked systems, is simulated via numerical integration of the full system of equations (see Supplementary Information). It is not due to a hysteretic behaviour of the system. Its origin is likely related with the specific anharmonic nature of the self-pulsing dynamics. For $V_{AC}>0.007V_{pi}$ the modulation at excitation frequencies $f_{mod}$ close to $f_{OMO,0}$ becomes too strong and the initial phonon lasing dynamics are lost, leading to a driven harmonic oscillator with a Lorentzian-like amplitude response in those regions and to an overall response that goes beyond pure locking.\\

\begin{figure}[t]
  \centering\includegraphics[width=\columnwidth]{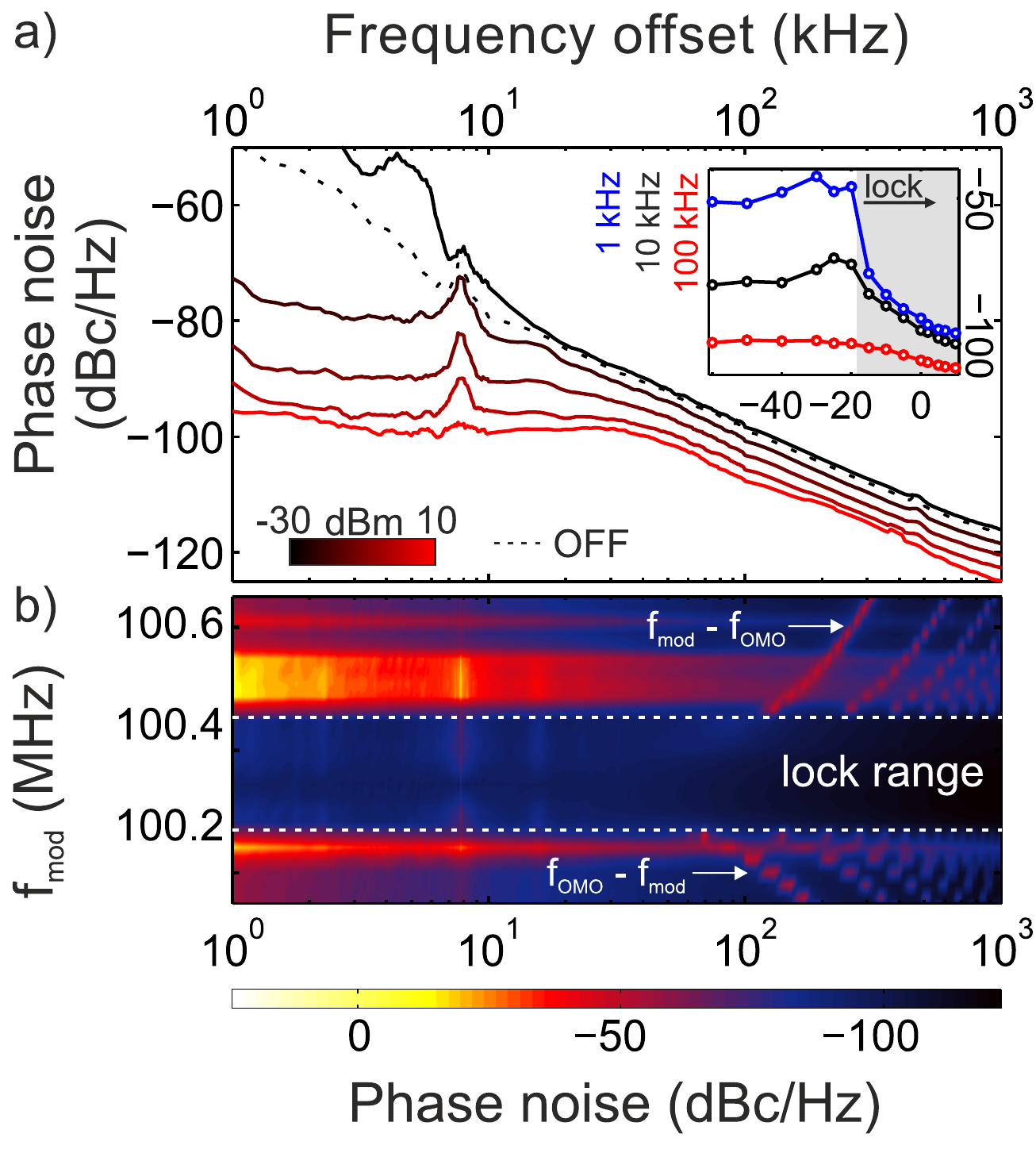}
    \caption{ \textbf{Phase noise characteristics.} (Color online) (a) Phase noise measurements of the phonon lasing state with a radio-frequency external tone at $f_{mod}\approx f_m$. The locked (unlocked) states are represented with black to red curves (dashed black line). The evolution of the phase noise at offsets 1, 10 and 100 kHz as a function of the modulation power $P_{VNA}$ are depicted in the inset, with the lock power range in shaded grey. (b) Color plot of the phase noise as a function of the frequency offset for multiple driving frequencies $f_{mod}$. The chosen range for $f_{mod}$ covers the full Arnold tongue for a drive power of $P_{VNA}=-10$ dBm.}
\label{fig:5}
\end{figure}

As seen in Fig.~\ref{fig:4}, injection-locking controls the oscillation frequency over bandwiths as high as several hundreds of linewidths of the bare OMO. Locking to a low-noise reference also improves its stability, gauged over short times via its phase noise. The phase noise analysis was carried out with a different fiber taper, slightly changing the external coupling condition and resulting in an even wider Arnold tongue at a fixed modulation amplitude and exhibiting injection-locking physics for voltages up to $V_{AC}=0.015V_{\pi}$ ($P_{VNA}=-10$ dBm). The phase noise measurements are carried out using the dedicated module of the SPA and setting the frequency carrier to the oscilator frequency $f_{OMO}$, which matches $f_{mod}$ inside the Arnold tongue. Fig.~\ref{fig:5}(a) depicts the phase noise as a function of frequency offset for different levels of modulation power, starting from the free-running phonon lasing (black dashed line). The curves are acquired with the modulation tone in close proximity to the off-modulation mechanical lasing frequency, i.e., $f_{mod}\approx f_{OMO,0}$, and checking the stability of the carrier power at $f_{OMO}$. We first see that for powers up to $P_{VNA}=-30$ dBm the phase noise increases for frequency offsets below 20 kHz. We attribute this to an unstable behaviour of the system, giving rise to hopping between the locked and the unlocked state due to environmental noise. At higher excitation powers the OMO locks to the external reference and, from that modulation amplitude, a strong and progressive phase noise reduction can be observed, especially at frequency offsets below 30 kHz but spanning the full range of the phase noise measurement setup up to 1 MHz. The broad peak around 7 kHz arises from an unknown noise source in the optical path and is disregarded. The phase noise level at frequency offsets of 1, 10 and 100 kHz is given as a function of modulation amplitude in the inset to Fig.~\ref{fig:5}. It shows both the increased low-offset phase noise prior to synchronization and an approximately linear phase noise reduction in the power lock range. A reduction of 1.1 dBc/Hz per excitation dBm is observed both at 1 kHz (blue) and 10 kHz (black), while the reduction is least pronounced at 100 kHz offset (red) and above. Given those values, the maximum achieved phase noise supression is 44 dBc/Hz (1 kHz), 20 dBc/Hz (10 kHz) and 8.5 dBc/Hz (100 kHz). Last, the behaviour of the phase noise at $P_{VNA}=-10$ dBm as a function of the modulation frequency $f_{mod}$ across the full Arnold tongue is assesed. When the modulation tone is swept from 100 MHz, outside the lock range, the modulation tone appears as a noise peak at the frequency difference (see Fig.~\ref{fig:5}(b)), as do all sidebands at frequency offsets $f=n\lvert f_{OMO}-f_{mod}\rvert$. Close to the onset of injection locking, at 100.22 MHz, where these additional peaks disappear, a considerable increase in the low frequency offset phase noise is observed, that is also attributed to hopping between the locked and unlocked state~\cite{pikovsky}. These considerations also apply in the opposite side of the lock range, where the increased phase noise is both more pronounced and extends throughout a broader range. Inside the Arnold tongue, the phase noise characteristics remain practically constant, evidencing the potential of injection-locking to generate low-noise oscillatory references and coherent phonon sources with controlled frequency and purity.\\

In conclusion, we have reported the observation of injection locking of self-pulsing-driven regenerative oscillations of an optomechanical system to an external optical reference and isolated the mechanical signal by engineering the system geometry and read-out scheme. The lock range achieved can go up to 350 kHz, enabling a large tuning range over the oscillator frequency. Besides the direct consequences for acoustic wave self-sustained sources of the results shown here, applications using distributed arrays of oscillators within a single small-footprint chip, like robust time-keeping with standard optoelectronic oscillators~\cite{timekeeping}, or neuromorphic computing\cite{neuromorphic}, may benefit from the reported physics of injection locking from a variety of perspectives. Variations in natural mechanical frequencies due to unwanted fabrication variations make mutual synchronization of weakly coupled beams as in Ref~\cite{synchro} relatively hard to observe. Due to the wide lock range achieved, injection locking can be used to pre-synchronize one of the coupled OMOs to an external drive and use it as a lead oscillator to which the others can synchronize via weak mechanical links, a configuration under exploration. Moreover, in cases where mechanical lasing of a particular mode cannot be achieved with an adiabatic laser parameter sweep, usually because the attained self-pulsing frequency $\nu_{SP}$ is still far from $f_m$, injection locking of the SP limit cycle to the external tone can be used to increase $\nu_{SP}$. With this additional degree of freedom mechanical lasing can be reached, leading to a state that persists after the modulation is switched off, i.e., it enables exploration of usually inaccessible dynamical attractors of the system. Pulsed or modulated operation in this configuration could enable switching between different dynamical states, as reported in~\cite{switching} with a top pumping scheme. Furthermore, the feedforward nature of the technique stabilizes the oscillator without the need of feedback circuits, achieveing a phase noise suppression up to 44 dBc/Hz at 1 kHz and 8.5 dBc/Hz at 100 kHz, resulting in an oscillator operated at atmospheric conditions with phase noise levels close to state of the art silicon optomechanical oscillators~\cite{aplwong,chiweiwong,laurita}.


\begin{widetext}

\newpage

\renewcommand{\figurename}{\textbf{Supplementary Figure}}
\makeatletter
\renewcommand{\thefigure}{S\@arabic\c@figure}
\makeatother
\renewcommand\theequation{S\arabic{equation}}
\renewcommand\thetable{S\arabic{table}}
\renewcommand{\bibname}{References}

\section{Supplementary information}

\setcounter{figure}{0}

\section{Self-pulsing mechanism.}
\label{sec:SPmec}

Driven optical micro and nano-cavities in the classical regime are often modelled using temporal coupled-mode theory, the specific formulation of which depends on the boundary conditions, i.e., on the way the cavity is excited~\cite{CPT}. For a photonic crystal cavity bidirectionally coupled to a bus waveguide, the fibre tapered loop in our experiments, used for excitation at power $P_{in}=\hbar\omega_L\lvert\overline{a}_{in}\rvert^2$ and frequency $\omega_L$, the equation governing the complex electric field amplitude (in the frame rotating at $\omega_L$) can be written as:
\begin{equation}
\label{CPTeq}
\frac{da(t)}{dt} = i\Delta a(t) - \frac{\kappa}{2}a(t) + \sqrt{\frac{\kappa_{e}}{2}}\overline{a}_{in}
\end{equation}
where $a(t)$ is normalized such that the intracavity photon number is given by $n_{c}(t)=a^*(t)a(t)$, $\Delta = \omega_{o}-\omega_{L}$ is the detuning, $\kappa$ the overall photonic decay rate and $\kappa_{e}$ the total photon decay rate into the extrinsic channel used for excitation, in this case the bus waveguide. The outcoupled field satisfies $a_{out}=a_{in}+  \sqrt{\frac{\kappa_{e}}{2}}a$, where we have considered that the bidirectional coupling is symmetric. This equation admits a simple steady state solution $\overline{a}$ for a CW-laser drive, the intra-cavity photon number $\overline{n}_{c}=\lvert\overline{a}\rvert^2$ being:
\begin{equation}
\label{eq:numberphotons}
\overline{n}_c=\frac{\kappa_e/2}{(\kappa/2)^2+\Delta^2}\frac{P_{in}}{\hbar\omega_L}
\end{equation}

The resonant circulating intensity, $I$, can be estimated~\cite{Vahala} in a high-$Q$ and small volume $V$ optical resonator as $I = P_{in} \left( \frac{\lambda}{2 \pi n_g}  \right) (Q/V)$, where $P_{in}$ is the input excitation power at wavelength $\lambda_L$ and $n_g$ is the group index of the resonant mode.\ For an input power of 1 mW and a typical silicon OMC mode with $Q \sim 5\cdot10^4$ and $V=2.4(\lambda/n)^3$, this corresponds to $1.8\,GW/m^2$.\ This large stored energy results in a significant non-linear behaviour of the dielectric medium. In the case of silicon, due to its central symmetry, the second order susceptibility tensor is null and only third order terms need to be considered~\cite{siliconsecondorder}. To such order, the main non-linear processes in silicon for single frequency operation in the telecomm range are two-photon absorption (TPA) and a dispersive Kerr effect, arising from the real and imaginary parts of the third order suspceptibility of electronic origin. Nevertheless, free carrier absorption (FCA) needs to be considered since a large population of free carriers $N_{e}$ can be generated, precisely due to two-photon absorption. A schematic describing the microscopic nature of these two phenomena can be found at the top of Fig.~\ref{fig:SP}. Last, most of the absorbed optical power in the cavity will be released to the lattice through the decay of the photoexcited carriers, increasing its temperature. The decay being much faster than the dynamics of the temperature, the energy transfer from the electron population to the lattice can be considered inmediate for the purposes of modelling the temperature field in the cavity. Such temperature rise in the cavity region will in turn produce a shift in the cavity resonant frequency $\omega_{o}$ of opposite sign to the one mediated by the presence of free carriers. Following the derivations in ~\cite{nonlinear1,nonlinear2,nonlinear3}, all such non-linear processes can be microscopically introduced into Maxwell's equations with the corresponding non-linear polarization terms, and then cast into the coupled mode formalism by writing:
\begin{subequations}
\begin{gather}
\frac{da(t)}{dt}= i\Delta a(t) - \left(\frac{\kappa}{2} + \frac{c^2}{n_{Si}^2}\frac{\hbar\omega_l\beta_{TPA}\lvert a(t)\rvert^2}{2V_{TPA}}+ \frac{c}{n_{Si}}\frac{\sigma_{r}N_{e}}{2 V_{FCA}} \right)a(t) + \sqrt{\kappa_{e}}\overline{a}_{in}  \label{eq:CPTnonlinear1}\\
\Delta= \omega_{L} - \left(\omega_{o}-\frac{\omega_{o}}{n_{Si}}\frac{\sigma_{i}N_{e}}{V_{FCA}}+\frac{\omega_{o}}{n_{Si}}n_{T}\Delta T\right) \label{eq:CPTnonlinear2}
\end{gather}
\end{subequations}
where, in addition to the cavity losses in the linear regime ($\kappa/2$), the absorbed power due to two-photon absorption and free carrier absorption have been considered in (\ref{eq:CPTnonlinear1}). Dispersion due to free carriers $N_{e}$ and temperature increase $\Delta T$ have also been introduced as can be seen in the equation for the detuning in (\ref{eq:CPTnonlinear2}). Here $\beta_{TPA}$ is the tabulated two-photon absorption coefficient~\cite{betaTPA}, $V_{TPA}$ and $V_{FCA}$ the characteristic volumes of the two-photon and free-carrier absorption processes, respectively, $n_{Si}$ the refractive index of silicon, c the speed of light, $\sigma_r$ and $\sigma_i$ the free-carrier absorption and dispersion \textit{cross-sections}~\cite{sigma} and $n_T$ the first-order refractive index variation caused by temperature. Here, we have already dismissed the dispersion associated to the Kerr effect due to the difference in magnitude compared to free-carrier dispersion and thermo-optic dispersion. We note here that linear absorption does not appear explicitly in (\ref{eq:CPTnonlinear1}) since its effect is already taken into account through the intrinsic cavity losses $\kappa_i$, that contribute to $\kappa$. This implies that solely considering linear absorption already requires solving a much more complex system in which $N_e$ and $\Delta T$ have a prominent role in the dynamics. Actually, in the sample explored in this manuscript, linear absorption dominates over the rest of the terms for the power levels considered and little to no difference is observed when disregarding the two-photon absorption term. The influence of the free carrier population $N_e$ and the temperature increase $\Delta T$ in Eqs. (\ref{eq:CPTnonlinear1}) and (\ref{eq:CPTnonlinear2}) imply that their dynamics need to be tracked too. These are obviously very complex space and time-dependent processes, but can be cast in the following form when the absorption processes mentioned and phenomenological decay rates are considered:
\begin{equation}\label{eq:FCD}
\frac{dN_{e}(t)}{dt}= -\gamma_{fc} N_{e} + \frac{1}{2}\frac{c^2}{n_{Si}^2}\frac{\hbar\omega_l\beta_{TPA}}{V_{TPA}}\lvert a(t)\rvert^4+\frac{c}{n_{Si}}\frac{\alpha_{lin}}{R_{eff}}\lvert a(t)\rvert^2
\end{equation}
\begin{equation}\label{eq:TO}
\frac{d\Delta T(t)}{dt}= -\gamma_{th} \Delta T + \frac{1}{\rho_{Si}C_{p,Si}V_{eff,T}} \left(\frac{c^2}{n_{Si}^2}\frac{\beta_{TPA}}{V_{TPA}}\lvert a(t)\rvert^4+\frac{c}{n_{Si}}\frac{\sigma_{r}N_{e}(t)}{V_{FCA}}\lvert a(t)\rvert^2+\frac{c}{n_{Si}}\frac{\alpha_{lin}}{R_{eff}}\lvert a(t)\rvert^2\right)
\end{equation}
where we have considered unavoidable linear absorption in addition to TPA and FCA. Here $\alpha_{lin}$ is the linear absorption coefficient~\cite{alphalin}, $R_{eff}$  represents the inverse of the fraction of the optical mode inside the silicon, $\rho_{Si}$ and $C_{p,Si}$ the density and constant-pressure specific heat capacity of silicon, $V_{eff,T}$ an effective thermal volume for the cavity and $\gamma_{fc}$ and $\gamma_{th}$ the free-carrier and thermal decay rates.\\

\begin{figure}[t]
\centering
\includegraphics[width=0.85\textwidth]{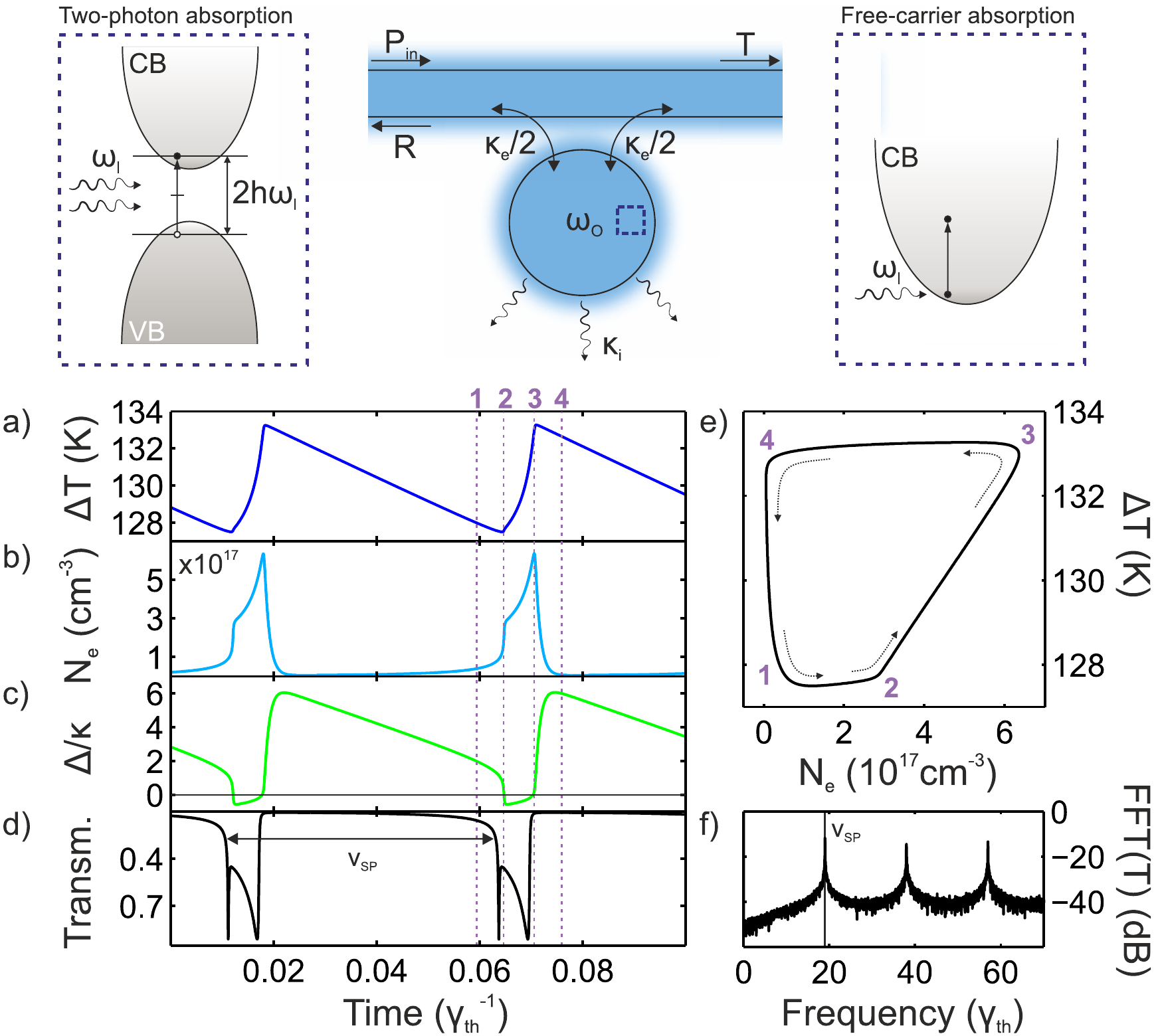}
\caption[radiationpressure]{\textbf{Self-pulsing dynamics of a highly driven silicon optical cavity}. Driven at \emph{high} powers, the dynamics of free-carriers $N_e$ and temperature changes $\Delta T$ need to be tracked. They couple dispersively to the optical resonance and can end up leading to periodic but highly anharmonic dynamical states. Both dynamical variables $\Delta T$ (a) and $N_e$ (b) display a periodic behaviour at frequency $\nu_{SP}$, forming a closed trajectory in phase space (e). The resulting time-periodic detuning $\Delta$ between the optical cavity and the driving laser modualtes the photon number $\overline{n}_c$ and the transmission, the typical time trace of which is shown in (d). (e) Fast Fourier transform (FFT) of the time trace. Time and frequency are given relative to the thermal decay rate $\gamma_{th}$.}
\label{fig:SP}
\end{figure}

An understanding of the dynamic behaviour of an optical cavity driven at high enough power requires the simultaneous solution of Equations (\ref{eq:CPTnonlinear1}), (\ref{eq:FCD}) and (\ref{eq:TO}). In order to simplify those, we can use the characteristic times involved in the different physical processes. For the optical modes we deal with in this work, the decay time of the optical mode ($\sim \text{ps}$) is orders of magnitude faster than the nonlinear dispersion mechanisms ($\sim \text{ns}$). As a consequence, we can asume $N_{e}$ and $\Delta T$ to be constant in the time-scales involved in (\ref{eq:CPTnonlinear1}). In addition, the losses induced by the non-linear processes can be first neglected with respect to the linear ones due to the apparent high presence of surface states in our SOI-based photonic structures, a possible reason for the low quality factors experimentally observed. Solving for the steady state intra-cavity photon number $\overline{n}_c$ as in (\ref{eq:numberphotons}), we can actually restrict our system of equations to (\ref{eq:FCD}) and (\ref{eq:TO}) by considering the adiabatic response of the optical cavity using the steady state number of photons (\ref{eq:numberphotons}) with a time-dependent detuning (\ref{eq:CPTnonlinear2}) and replacing $\rvert a(t)\lvert^2=\overline{n}_c(t)$ in (\ref{eq:FCD}) and (\ref{eq:TO}). These simplified equations can be integrated without much computational effort and capture most of the experimental features in the presented experiment as well as in our previous experiments~\cite{dani_selfpulsing}. We consider all parameteres defining the equations above, except the laser parameters $P_{in}$ and $\omega_L$, as given. In most situations, the dynamic solution to the system of differential equations is a fixed or equilibrium point in the phase space defined by $\{N_{e},\Delta T\}$ which leads to a stable spectral shift of the cavity mode, typically dominated by the thermal part and referred to as a thermo-optic shift~\cite{TOeffect}. For particular combinations of $P_{in}$ and $\omega_L$ stable limit cycles exist in phase space. With initial conditions $\{N_{e}(0),\Delta T(0)\}$ inside the bassin of attraction of this limit cycle, the dynamic solution in the limit $t\to\infty$ tends to the cycle, forming a periodic closed trajectory in phase space $\{N_{e}^*(t),\Delta T^*(t)\}$ - a self-pulsing (SP) limit cycle~\cite{nonlinear1}- which is generally highly anharmonic. The typical shape of this closed trajectory, as well as the resulting temporal traces for $N_{e}$, $\Delta T$, $\Delta$ and the transmission $T$ are obtained by numerical integration and are depicted in Fig.~\ref{fig:SP}(a,b,c,d,e). Panel (f) sows the Fast Fourier Tranform (FFT) of the transmission trace in (d), taken over many oscillation periods.\\

\begin{figure}[t]
\centering
\includegraphics[scale=0.53]{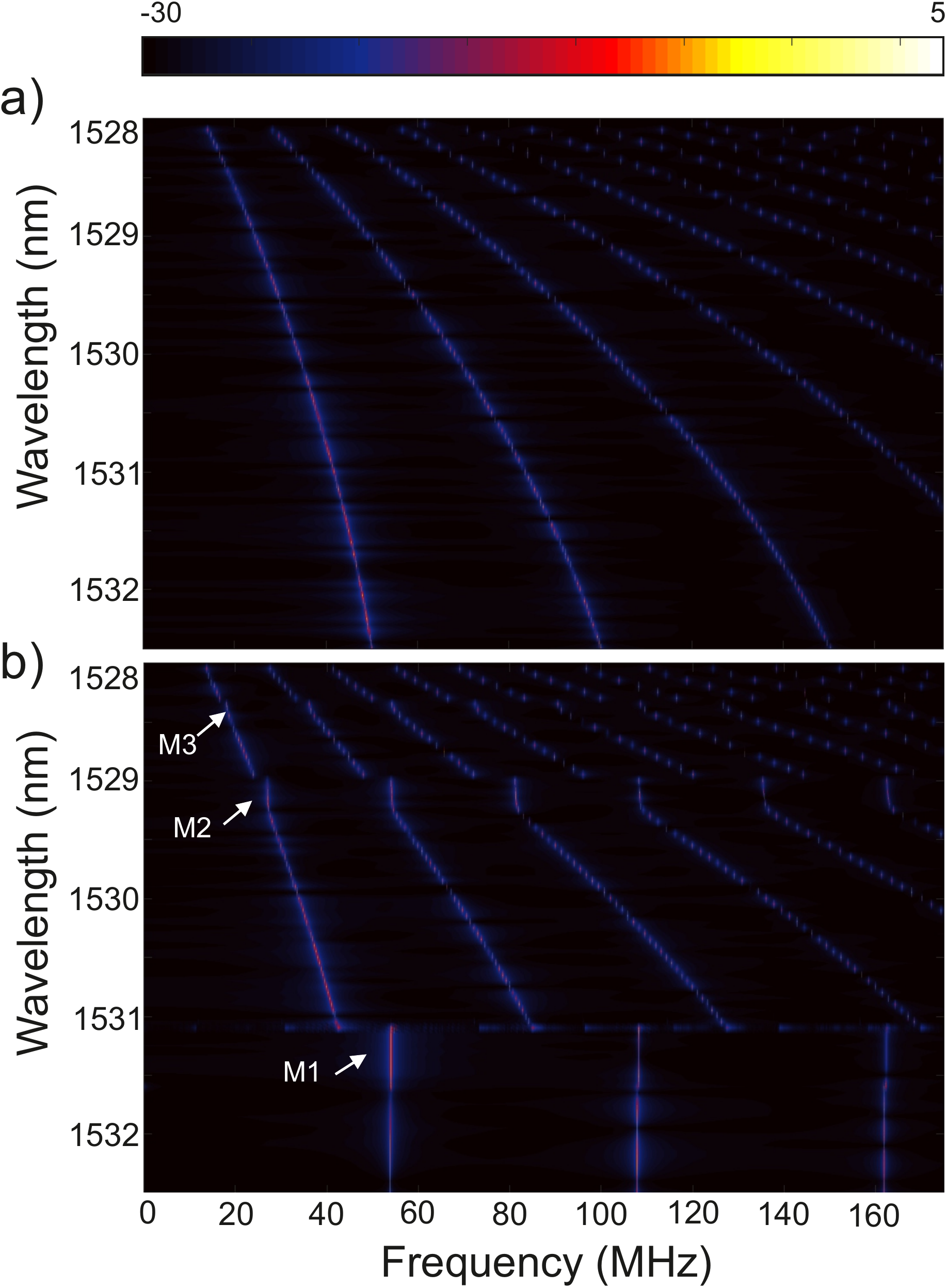}
\caption[Simulated radiofrequency spectra as a function of wavelength with and without the mechanical resonator]{\textbf{Simulated radiofrequency spectra as a function of wavelength with and without the mechanical resonator.} (a) Radiofrequency spectra of the transmission time trace as a function of the laser wavelength $\lambda_L$ without (a) and with (b) the mechanical mode at $\Omega_m=54$ MHz. When the mechanical motion is taken into account, the system can evolve to mechanical lasing, driven by the self pulsing-induced modulation of the cavity photons $n_c(t)$. Frequency entrainment occurs when $\nu_{SP}$ is close to being an integer multiple of the mechanical frequency $m \nu_{SP}=\Omega_m$. Three such values of $m$ are highlighted with arrows and are denoted as M3, M2 and M1, depening on the value of $m$.}
\label{fig:SPandMEClambda}
\end{figure}

Since in most experiments our knob is the laser wavelength $\lambda_L$ (the laser frequency $\omega_L$) it is interesting to study how the SP evolves with $\lambda_L$ as this one is scanned accross the optical cavity mode of frequency $\omega_{o}$. Numerically solving the set of equations described above for many laser wavelengths leads to the colormap of the radiofrequency spectrum shown in Fig. \ref{fig:SPandMEClambda}(a). To mimick experimental measurement conditions, these simulations start with a laser drive on the blue side of the resonance. The initial conditions at for numerical integration at each wavelength correspond to the the solution $\{N_{e}(t_*),\Delta T(t_*),x(t_*)\}$ of the previous wavelength, with $t_*$ chosen such that the solution has reached a steady-state.\\

\section{Mechanical lasing.}
\label{sec:meclasing}

The system of equations described above only considers the optical response of the system. If the optical cavity of interest is coupled to a mechanical mode of frequency $\Omega_m$ and dissipation rate $\Gamma_m$ via radiation-pressure with a vaccuum optomechanical coupling rate $g_o$, the dynamics of the mechanical mode might have to be considered, at least for particular values of the laser drive $P_{in}$ and $\lambda_L$. Even if the coupling rate $g_o$ is low or if the system is far from the sideband resolved condition ($\kappa<\Omega_m$), where the response of the mechanical resonator to an external stimulus is truly modified~\cite{cavityoptomechanics}, the periodic modulation of the photon number $\overline{n}_c(t)$  at frequency $\nu_{SP}$ induced by the development of a SP limit-cycle can drive the mechanical mode into a coherent oscillation state. This happens whenever $\nu_{SP}$ or any of its harmonics $M$ is close to the mechanical frequency $\Omega_m$. Since the SP mechanism has no fundamental natural frequency of its own, once the mechanical resonator is excited the self-pulsing is entrained  to oscillate at the frequency of the mechanical mode for some range of the drive parameters. Actually, for considerably large absolute values of $g_o$ and high mechanical dissipation $\Gamma_m$ the self-pulsing might never occur without excitation of the mechanical mode, leading to sharp jumps between lasing states with different harmonics $M$ of the SP limit cycle, which is what we observe in our experiment and is depicted in Figure 1 of the main text. To understand the observed dynamics, the following system of equations now needs to be solved:

\begin{equation}\label{eq:FCD2}
\frac{dN_{e}(t)}{dt}= -\gamma_{fc} N_{e} + \frac{1}{2}\frac{c^2}{n_{Si}^2}\frac{\hbar\omega_l\beta_{TPA}}{V_{TPA}}\overline{n}_c(t)^2+\frac{c}{n_{Si}}\frac{\alpha_{lin}}{R_{eff}}\overline{n}_c(t)
\end{equation}
\begin{equation}\label{eq:TO2}
\frac{d\Delta T(t)}{dt}= -\gamma_{th} \Delta T + \frac{1}{\rho_{Si}C_{p,Si}V_{eff,T}} \left(\frac{c^2}{n_{Si}^2}\frac{\beta_{TPA}}{V_{TPA}}\overline{n}_c(t)^2+\frac{c}{n_{Si}}\frac{\sigma_{r}N_{e}(t)}{V_{FCA}}\overline{n_c}(t)+\frac{c}{n_{Si}}\frac{\alpha_{lin}}{R_{eff}}\overline{n}_c(t)\right)
\end{equation}
\begin{equation} \label{eq:nonlinearmechanics}
\begin{aligned}
\frac{d^2 x(t)}{dt^2}+\Gamma_m\frac{dx(t)}{dt}+\Omega_m^2x(t)=\hbar \frac{g_o}{x_{zpf}m_{eff}}\overline{n}_c(t)
\end{aligned}
\end{equation}
where $x(t)$ is the generalized coordinate for the displacement of the mechanical mode and $m_{eff}$ its effective mass. The two first equations correspond to the ones solved in the previous Section and the third describes the dynamics of the mechanical resonator under the action of radiation pressure force $F_{RP}(t)=\hbar\frac{g_o}{x_{zpf}}n(t)$. In this last equation, the action of thermal Langevin forces is left out for simplicity, although those are necessary to experimentally observe the mechanical modes at low driving powers. Due to the low frequency MHz modes considered herein, these three equations are solved again considering the adiabatic response of the optical cavity to the rest of the dynamics, i.e. $n_c(t)=\overline{n}_c(t)$ which implies
\begin{equation}
\label{eq:numberphotonstime}
\overline{n}_c=\frac{\kappa_e/2}{(\kappa/2)^2+\Delta^2}\frac{P_{in}}{\hbar\omega_L}
\end{equation}
and the detuning
\begin{equation}
\label{eq:detuntotal}
\Delta= \omega_{l} - \left(\omega_{o}-\frac{\omega_{o}}{n_{Si}}\frac{\sigma_{i}N_{e}}{V_{FCA}}+\frac{\omega_{o}}{n_{Si}}n_{T}\Delta T-\frac{g_o}{x_{zpf}}x\right) 
\end{equation}
now includes the dispersive effect of the mechanical mode displacement on the optical cavity.\\

For ease of understanding, Fig.~\ref{fig:SPandMEClambda}(b) depicts what happens for an optomechanical system where both regions of pure self-pulsing and mechanical lasing are visible. Several entrainment plateaus show the different orders $M$ of the mechanical lasing, where the $M$-th harmonic of the SP frequency, $M\nu_{SP}$ coincides with the mechanical frequency $\Omega_m$. Except for the inclusion of the mechanical dynamics and the optomechanical coupling rate $g_o$, the rest of the parameters are kept as in the previous section. This allows us to understand how the SP mechanism acts as a driving term for the mechanical lasing, as can be easily seen by comparing Figures \ref{fig:SPandMEClambda}(a) and (b), where regions of pure SP are practically equivalent. Whenver $g_o$ or $\Gamma_m$ increase, it is easier to drive the mechanical mode at a frequency difference $\Omega_m-n\nu_{SP}$, and a minimal drive can end up locking the SP to a subharmonic, which results in spectra similar to the one observed experimentally, where regions of pure SP are not observed.\\

\section{Injection-locking dynamics.}

We develop now a numerical model to qualitatively describe the experimental results obtained for the injection locking experiment. For such purpose, we only have to add the effect of modulating the incoming laser power $P_{in}$ via the electro-optic modulator (EOM). The time-dependent input power is given by $P_{mod}=P_{in}\text{cos}^2\left( \pi \frac{V-V_{\pi}}{2V_{\pi}}\right)$, where $P_{in}$ is again the input excitation power entering the EOM, $V$ the voltage applied to the EOM and $V_{\pi}$ is the characteristic half-wave voltage, which for our EOM is $V_{\pi}=6.7 V$. The voltage used in this work is given by $V=V_{DC}+V_{AC}\text{cos}(2\pi f_{mod}t)$, where $V_{DC}$ is fixed at the optimal modulation point $V_{DC}=V_{\pi}/2$ and $V_{AC}$ can be varied by using the VNA as an RF source. \\

\begin{figure}[t]
\centering
\includegraphics[scale=0.9]{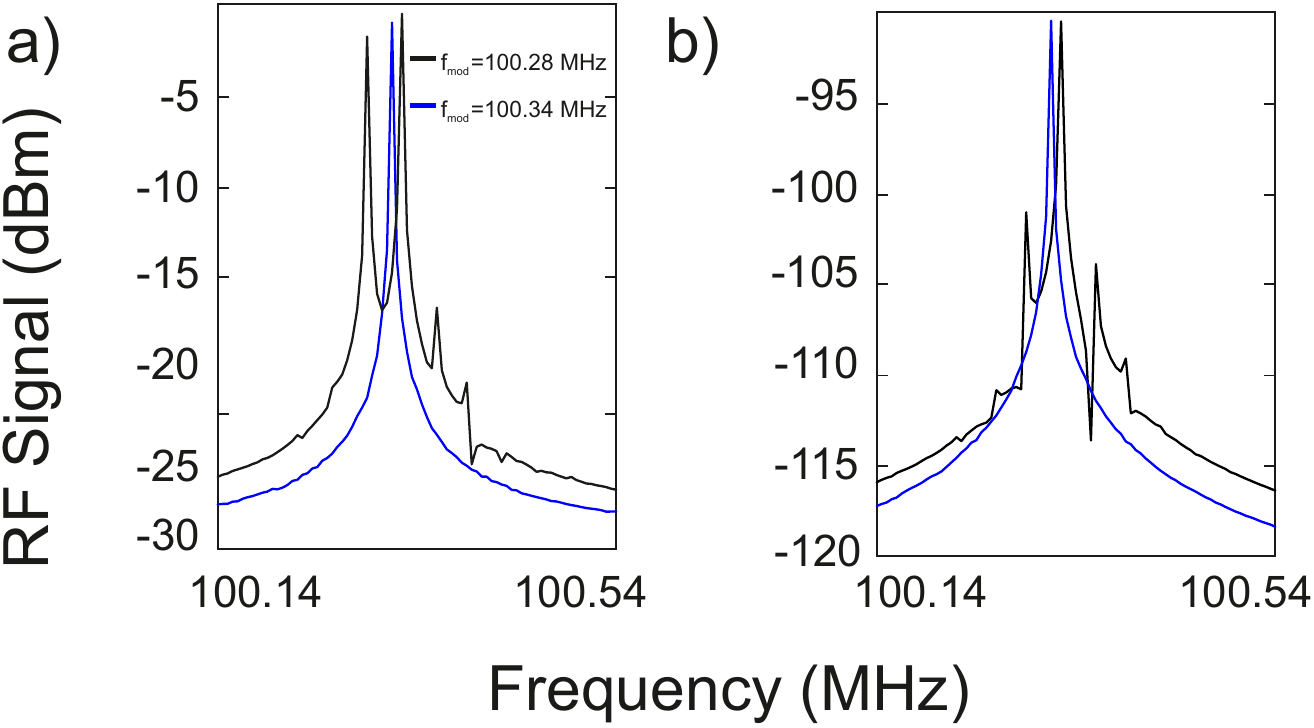}
\caption[radiationpressure]{\textbf{Radiofrequency spectra in the injection locking regime} Power spectral density (PSD) of the (a) transmission and (b) mechanical time traces. An external modulation frequency far from the lasing frequency ($f_{mod}=$ 100.28 MHz, black solid line), leads to two different peaks, while the external tone locks the mechanical lasing frequency for small differences ($f_{mod}=$ 100.34 MHz, blue solid line).}
\label{fig:num_sim1}
\end{figure}

\begin{figure}[t!]
\centering
\includegraphics[width=0.8\textwidth]{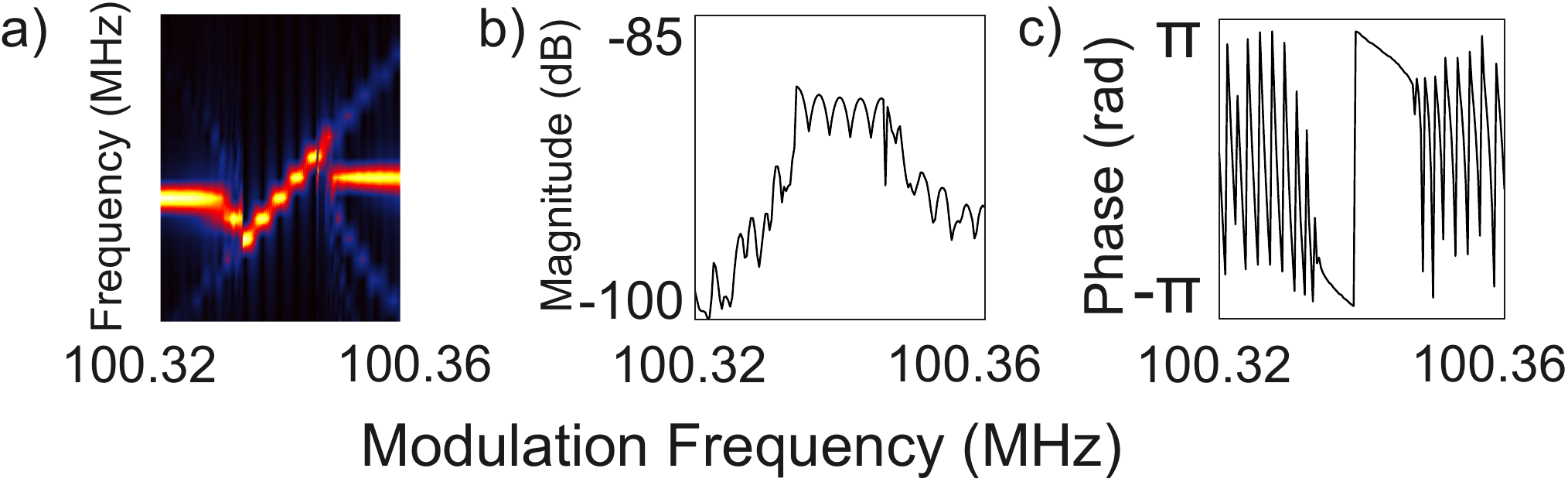}
\caption[radiationpressure]{\textbf{Numerical results reproducing the coherent response of the mechanical signal}  (a) Color plot of the PSD of the transmitted signal obtained from the Right cavity for a modulation voltage of $V_{AC}$= 0.034 $V_{\pi}$. (b) (c) Coherent magnitude and phase response obtained from the temporal traces of the Right cavity.}
\label{fig:num_sim2}
\end{figure}

Using the same time-scale arguments as the ones evoqued for the free-carrier, temperature and mechanical dynamics, the response of the optical cavity to the power modulation can be considered instantaneous (the modulation frequencies $f_{mod}$ are close to the mechanical frequency $\Omega_m$)  and we can this time set the intracavity photon number shown in Eq. (\ref{eq:numberphotonstime}) as:
\begin{equation}
\label{eq:numberphotonsmod}
\overline{n}_c=\frac{\kappa_e/2}{(\kappa/2)^2+\Delta^2}\frac{P_{mod}}{\hbar\omega_L}
\end{equation}
which implies that the numerical resolution of the equations is practically equivalent to what is done for the standard mechanical lasing. Note however that being able to spectrally resolve the two peaks, the modulation and lasing peaks, whenever they are very close in frequency requires extremely long simulation time windows. For example, for  a frequency difference $ \lvert\Omega_{OMO}-f_{mod}\rvert=0.01$ MHz in the unlocked regime, resolving the two peaks requires a time window of (at the very) least 100 $\mu s$. However, even longer times are required in practice since the solution only becomes stable after some initial time. The simulations shown here use a time span $t\in[0,400\mu s]$.\\

\begin{figure}[b]
\centering
\includegraphics[width=0.75\textwidth]{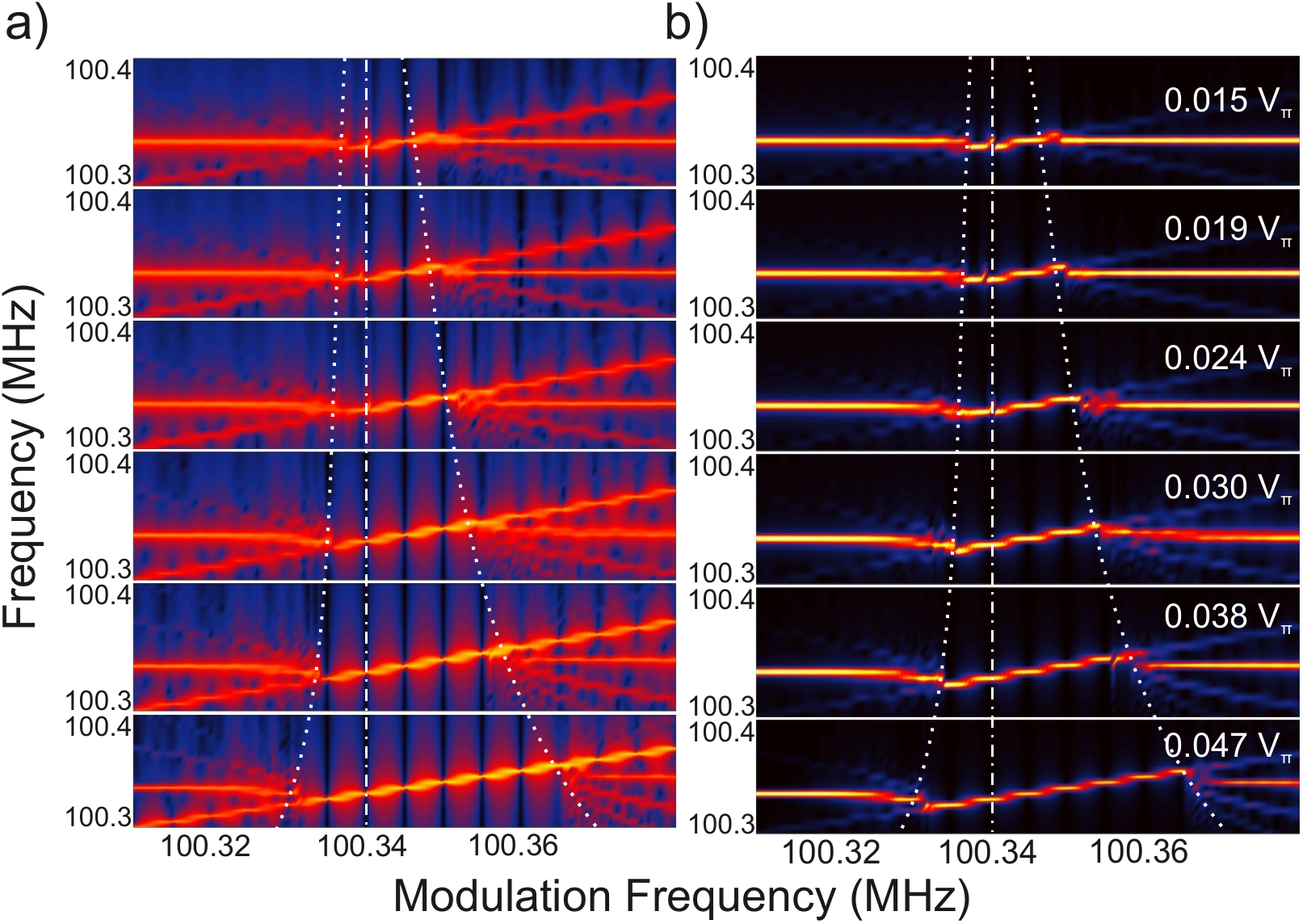}
\caption[Simulated Arnold tongue]{\textbf{Simulated Arnold tongue as a function of the modulation depth $V_{AC}$}. (a) Color plot of the transmitted signal FFT spectra and (b) the mechanical displacement FFT spectra from $V_{AC}= 0.0149 V_{\pi}$ to $V_{AC}= 0.0472 V_{\pi}$. The vertical dashed white line highlights the oscillator frequency in the absence of any drive $\Omega_{OMO,0}$, while the other two delimit the Lock range and evidence an asymmetric growth on both sides of $\Omega_{OMO,0}$. }
\label{fig:Arnold}
\end{figure}

As explained in the main text, the use of a neighbour optical cavity allows us to readout the mechanical dynamics via the non-vanishing cross coupling terms. In any case, what we can be certain is that the thermal and free-carrier effects are extremely low when reading out the temporal dynamics of the outcoupled light in the neighbour cavity. This implies that whatever we observe in that cavity can be directly compared to what is numerically obtained for the mechanical resonator dynamics. Fig.~\ref{fig:num_sim1} shows the numerical solutions obtained for the injection locking scheme. Fig. \ref{fig:num_sim1}(a) and (b) show the PSD obtained from both the transmission and mechanical displacement time traces, mimicking the spectrum obtained while measuring on the right (R) and left (L) respectively. The black curves correspond to the spectrum for a modulation frequency $f_{mod}$ far from the mechanical frequency. The absence of direct modulation on the mechanical time trace implies that the peak at the modulation frequency is obviously much smaller than the mechanical lasing peak. The blue plots correspond to a small detuning between the excitation frequency and the mechanical frequency, and we clearly see that only a single peak is observed, which is what correspond to the injection locked regime. Fig. \ref{fig:num_sim2}(a) shows the color plot of the PSD signal as obtained from the mechanical displacement time traces, illustrating the lock range for a modulation voltage of $V_{AC}=0.034V_{\pi}$. Fig. \ref{fig:num_sim2}(b) and (c) show the reconstructed magnitude and phase response from the simulated temporal traces of the mechanical displacement. The recovered response agrees well with the experimental measurements shown in Figure 2(b) and (d) of the main manuscript.\\

The Arnold tongues shown are obtained by integrating the full system of equations for several modulation frequencies $f_{mod}$, using the solution $\{N_{e}(t_*),\Delta T(t_*),x(t_*)\}$ of the previous modulation frequency as an initial condition for each value of $f_{mod}$. To mimick the experimental observation, we show in Fig.~\ref{fig:Arnold} the Arnold tongue as observed from the PSD colormaps of both the transmission time traces (a) and the mechanical time traces (b). Much like in the experiment, whose spectra are shown in Figure 3 in the main text, the lock range grows asymmetrically on both sides of the mechanical frequency (white dashed line). The qualitative agreement with the experimetally obtained Arnold tongues is obvious, although the size of the lock ranges for a particular modulation amplitude $V_{AC}$ is much larger in the experimental setting. This discrepancy is likely due to the difficulty in having a precise value for all of the governing parameters in the non-linear dynamical equations of the optical cavity. \\

\section{Higher-harmonic injection-locking.}

\begin{figure}[t!]
\centering
\includegraphics[width=0.75\textwidth]{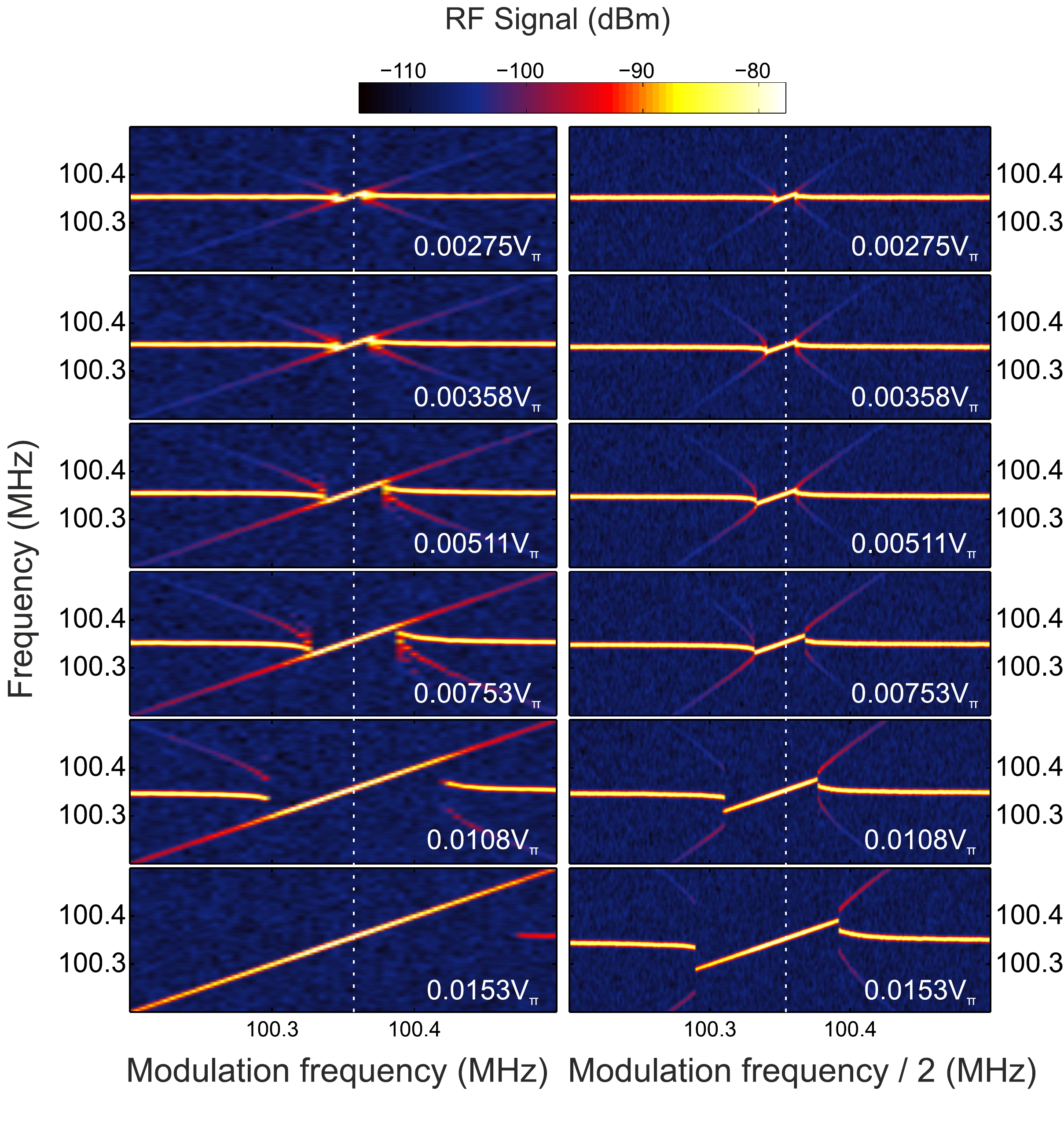}
\caption[Experimental Arnold tongue second harmonic]{\textbf{Experimental Arnold tongue at the first harmonic}.  (Color online) Colormaps of the power spectral density (PSD) of light transmitted at the wavelength of the probe laser for modulation amplitudes from $V_{AC}$ = 0.00275$V_{\pi}$ to $V_{AC}$ = 0.0153$V_{\pi}$. (Left) Modulation frequency $f_{mod}$ around $f_{OMO,0}$ and (right) modulation frequency $f_{mod}$ around $2f_{OMO,0}$. The dashed white lines mark either the frequency of the free-running oscillator at $f_{OMO,0}$=100.355 MHz or its first harmonic.}
\label{fig:n2sync}
\end{figure}

Injection-locking of a regenerative oscillator can also be achieved close to higher-order harmonics of the oscillation frequency, i.e., at $f_{mod}\sim nf_{OMO}$~\cite{pikovsky}. This was checked experimentally by carrying out measurements like the ones shown in Fig. 4 of the main text but using a modulation frequency around the first harmonic, i.e. n=2. This is shown, along with measurements done at the direct oscillation frequency, i.e. n=1, in Fig.~\ref{fig:n2sync}. We see that on this measurement run the asymmetry in the Arnold tongue for direct modulation at $f_{mod}\sim f_{OMO}$ does not show up, while the Arnold tongue when driving close to the first harmonic exhibits an asymmetry towards the opposite direction as the one shown in the main text. This is currently under investigation in a configuration in which we injection-lock only the SP dynamics.

\section{Mode hybdridization and phonon detection.}

\begin{figure}[t]
\centering
\includegraphics[width=\textwidth]{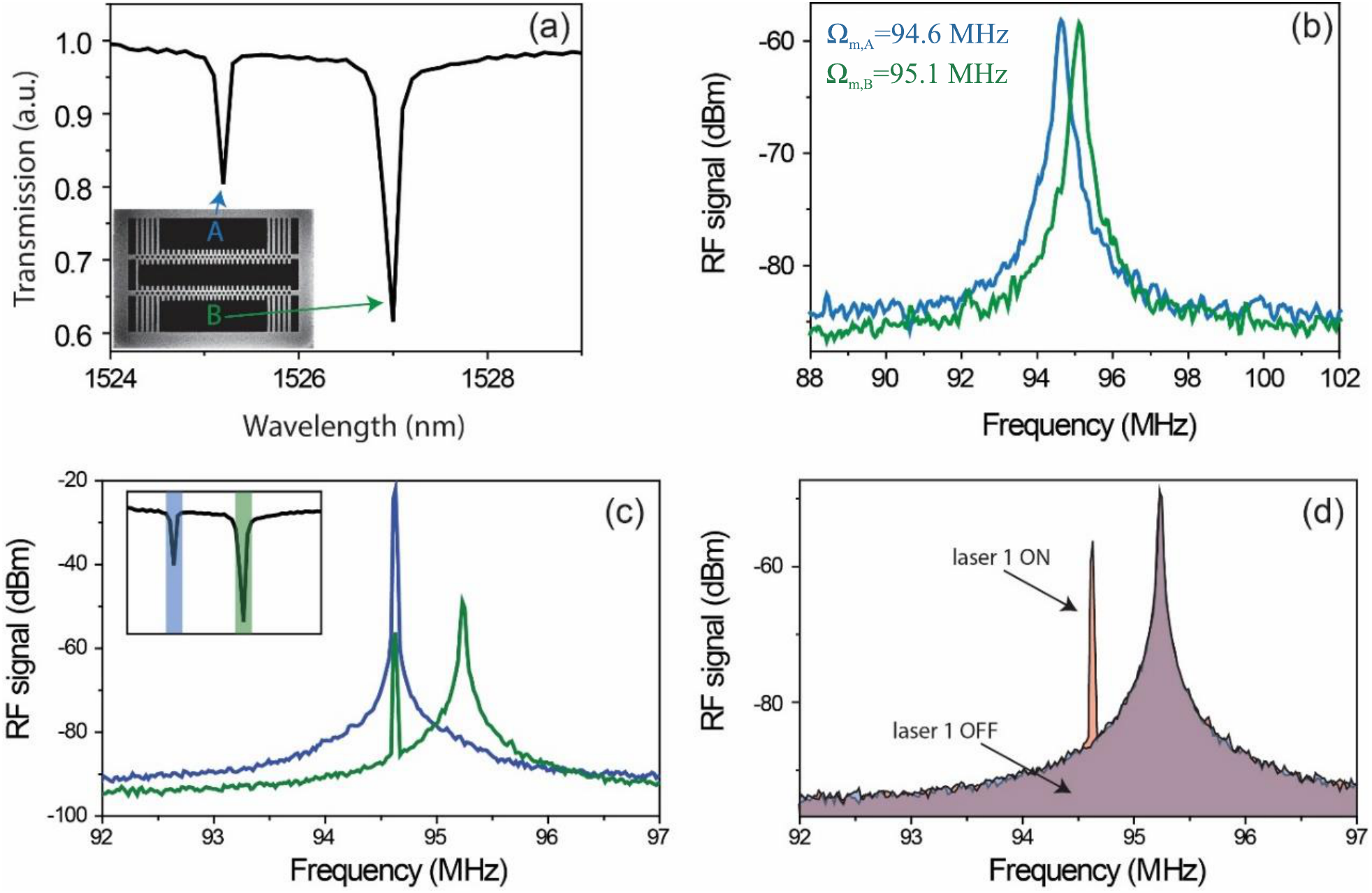}
\caption[radiationpressure]{\textbf{Phonon detection}. (a) Optical resonance belonging to cavity A and B when coupled through a common optical fiber. (b) Thermally driver mechanical resonances. The small detuning is due to fabrication imperfections. (c) Detection of the cavity A in mechanical lasing (blue) by employing cavity B (red), mechanically coupled via a common tether. (d) The detected signal follows the phonon laser switching.}
\label{fig:modeshybrid}
\end{figure}

As mentioned in the manuscript, the string-like mechanical modes of both beams would have exactly the same oscillation frequency $\Omega_{m,A}=\Omega_{m,B}$ in the perfect case and would hybridize into a symmetric (+) and antisymmetric (-) modes with frequencies $\Omega_{m,\pm}=\Omega_{m,A}\pm G$, with $G$ the coupling induced by the common tether. Fabrication disorder breaks the symmetry and each mode has a different original oscillation frequency $\Omega_{m,A}$ and $\Omega_{m,B}$. For our structures, the level of fabrication disorder is relatively strong and the uncoupled detuning $\Delta_m=\Omega_{m,A}$-$\Omega_{m,B}$ is considerable, much bigger than the $G$ induced by the coupling tether(s) for this type of string-like mode. This implies that, once coupled, $\Omega_{m,+}\sim\Omega_{m,A}$ and $\Omega_{m,-}\sim\Omega_{m,B}$ or viceversa. Assuming the former case and that all other normal modes of the uncoupled resonators are far in energy (which is obviously the case for this type of mode), the modes can be written like 
\begin{subequations}
\begin{align}
\lvert + \rangle_m &= \lvert A\rangle_m + \alpha_m\lvert B\rangle_m \\
\lvert - \rangle_m &= \lvert B\rangle_m - \alpha_m\lvert A\rangle_m
\end{align}
\end{subequations}
with $0<\alpha_m\ll 1$. The same happens for the optical modes of the structure but due to the stronger dependence of the mode frequencies on the existing fabrication disorder (hole and wing sizes, rugosity, loss of verticality, etc.) we can assume -and this is confirmed by the huge experimental detunings observed- that they do not couple at all and set $\alpha_o=0$. In such case, we define the optomechanical coupling rates $g_o$ between the original optical and mechanical modes as $g_{o,\lvert A\rangle_m\lvert A \rangle_o}=g_A$, $g_{o,\lvert B\rangle_m\lvert B \rangle_o}=g_B$, $g_{o,\lvert A\rangle_m\lvert B \rangle_o}=0$ and $g_{o,\lvert B\rangle_m\lvert A \rangle_o}=0$, where in principle we could expect $g_A$ to be similar to $g_B$. However, one can easily see that for the mechanical mode considered in this work, which is asymmetric with respect to the nanobeam axis, one would expect no coupling at all to any symmetric/antisymmetric optical mode $g_A=g_B=0$. However, as is obtained from direct numerical simulation of the geometries obtained from high-resolution SEM micrographs, the optical modes develop a modified shape due to asymmetries in the fabricated structure, leading to an effective, and in this particular example, high optomechanical couplings $g_{A}$ and $g_{B}$ which can be of different absolute value and sign, depending on where the modified field is located. Indeed, if the field relocates to one side of the beam, that side is contracted when the other side of the beam is expanded in a string-like mode, leading to different sign of $g_{A/B}$. When the tether is added, the optomechanical coupling terms can be cast like 
\begin{subequations}
\begin{align}
g_{o,\lvert +\rangle_m\lvert A \rangle_o}&=g_A \\
g_{o,\lvert -\rangle_m\lvert B \rangle_o}&=g_B \\
g_{o,\lvert +\rangle_m\lvert B \rangle_o}&=\alpha_m g_B \\
g_{o,\lvert -\rangle_m\lvert A \rangle_o}&=\alpha_m g_A  
\end{align}
\end{subequations}
which grants that, whatever the signs of $g_A$ and $g_B$, the read-out of a particular oscillating mechanical mode, e.g. $\lvert +\rangle_m$, with optical modes A and B leads to a particular phase relation, while the read-out of the other mode reads the same phase relation shifted by half a cycle.\\
\begin{figure}
\centering
\includegraphics[width=\textwidth]{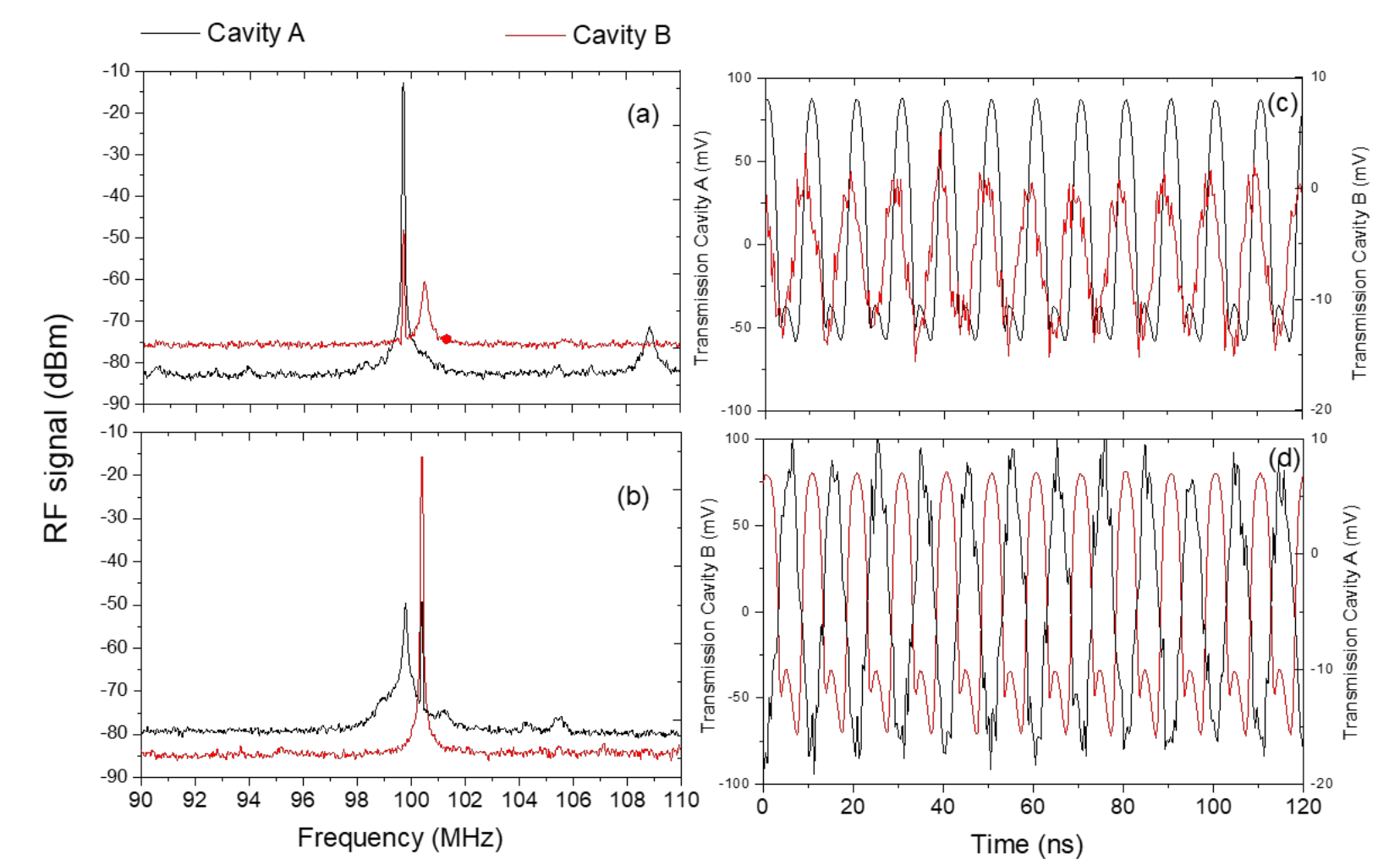}
\caption[radiationpressure]{\textbf{Activation and detection of either symmetric or antisymmetric mechanical modes.} RF spectra and temporal traces when Cavity A is in a lasing state and Cavity B is just transducing (panels a and c) and viceversa (panels b and d). Black (Red) curves correspond to traces recorded when detecting the transmitted signal coming from Cavity A (B).}
\label{fig:timetrace}
\end{figure}
The value of $\alpha_m$ is very weak in all of our experiments and therefore we need the mechanical oscillator A (B) to be in a high amplitude state to be able to experimentally observe any signal by driving the neighbour optical cavity B (A). Since the OMCs used for the experiments in the main text did not both reach the mechanical lasing state simultaneously, we present here results obtained from another pair of coupled beams, with very similar geometry, only with slightly longer structures, leading to lower mechanical frequencies $\Omega_A$ and $\Omega_B$ than those observed in the main text. When weakly driving cavity A (B) and transducing the thermal motion of mode A (B) we are unable to observe any signal above the noise floor of our spectrum analyzer coming from thermal motion of nanobeam B (A), as can be seen on Fig.~\ref{fig:modeshybrid}(b). In panel (c) of the same figure, we see that when nanobeam A (blue) is brought to a mechanical lasing state as the one described previously, we can also add a weak probe driving cavity B (green) and read out both the mechanics of cavity A and its own thermal motion. As seen in (d), the additional peak disappears when the laser driving optical mode A is turned off, evidencing the \emph{distant} phonon detection mechanism. This independent assessment is achieved by using two narrow bandpass filters (BPFs) as discussed in the main text. The experiment can be also reproduced in the reverse way (Fig.~\ref{fig:timetrace}(a,b)), allowing us to study the hybrid nature of the modes in more detail by looking at averaged temporal traces, where any signal coming from the existing thermal motion vanishes. We see (Fig.~\ref{fig:timetrace}(c)) that the signal of nanobeam A in a lasing state as read-out via its own optical mode and via optical mode B have a fixed phase relation and that the fixed phase relation obtained for the reverse case (d) is increased by $\pi$. As stated above, this confirms the presence of a phase and an anti-phase oscillation of both beams in the two mechanical eigenmodes, even though we cannot infer which mode is symmetric and which one is asymmetric without knowing the particular signs of the coupling terms.

 \end{widetext}


\begin{thebibliography}{99}

\bibitem{cavityoptomechanics}
M. Aspelmeyer, T. J. Kippenberg and F. Marquardt. Rev. Mod. Phys. \textbf{86}, 1391 (2014).

\bibitem{chan}
J. Chan, T. P. M. Alegre, A. H. Safavi-Naeini, J. T. Hill, A. Krause, S. Groeblacher, M. Aspelmeyer and O. Painter. Nature \textbf{478}, 89-92 (2011).

\bibitem{amplification}
T. J. Kippenberg, H. Rokhsari, T. Carmon, A. Scherer, and K. J. Vahala. Phys. Rev. Lett. \textbf{95}, 033901 (2005).

\bibitem{vahalaoscillator}
M. Hossein-Zadeh and K. J. Vahala. IEEE J. Sel. Top. Quantum Electron. \textbf{16}, 276 (2010).

\bibitem{electriccontrol1}
E. Baldini, T. Palmieri, A. Dominguez, P. Ruello, A. Rubio, and M. Chergui. Nano Letters \textbf{18}, 5007 (2018).

\bibitem{electriccontrol2}
A. Hern\'andez-M\'inguez, Y.-T. Liou, and P. V. Santos. J. Phys. D: Appl. Phys. \textbf{51}, 383001 (2018).

\bibitem{opticcontrol1}
A. S. Kuznetsov, K. Biermann, and P. V. Santos, Phys. Rev. Research \textbf{1}, 023030 (2019).

\bibitem{opticcontrol2}
T. Czerniuk, C. Br\"uggemann, J. Tepper, S. Brodbeck, C. Schneider, M. Kamp, S. H\"ofling, B. A. Glavin, D. R. Yakovlev, A. V. Akimov, and M. Bayer. Nature Communications \textbf{5}, 4038 (2014).

\bibitem{magneticcontrol1}
M. Weiler, H. Huebl, F. S. Goerg, F. D. Czeschka, R. Gross, and S. T. B. Goennenwein. Phys. Rev. Lett. \textbf{108}, 176601 (2012).

\bibitem{magneticcontrol2}
R. Mankowsky, A. von Hoegen, M. F\"orst, and A. Cavalleri. Phys. Rev. Lett. \textbf{118}, 197601 (2017).

\bibitem{painterwaveguide}
K. Fang, M. H. Matheny, X. Luan, and O. Painter. Nature Photonics \textbf{10}, 489 (2016).

\bibitem{phononsinfo}
W. Fu, Z. Shen, Y. Xu, C.-L. Zou, R. Cheng, X. Han and H. X. Tang. Nature Communications \textbf{10}, 2743 (2019).


\bibitem{phononlasing2}
T. Carmon, H. Rokhsari, L. Yang, T. J. Kippenberg and K. J. Vahala. Phys. Rev. Lett. \textbf{94}, 223902 (2005).

\bibitem{phononlasing3}
I. S. Grudinin, H. Lee, O. Painter and K. J. Vahala. Phys. Rev. Lett. \textbf{104}, 083901 (2010).


\bibitem{ghorbel}
I. Ghorbel, F. Swiadek, R. Zhu, D. Dolfi, G. Lehoucq, A. Martin, G. Moille, L. Morvan, R. Braive, S. Combri\'e and  A. De Rossi. APL Photonics \textbf{4}, 116103 (2019).

\bibitem{laurita}
L. Mercad\'e, L. L. Mart\'in, A. Griol, D. Navarro-Urrios and A. Mart\'inez. Nanophotonics \textbf{9}, 3535 (2020).

\bibitem{bolometric}
C. Metzger, M. Ludwig, C. Neuenhahn, A. Ortlieb, I. Favero, K. Karrai and F. Marquardt. Phys. Rev. Lett. \textbf{101}, 133903 (2008).

\bibitem{scireports}
D. Navarro-Urrios, N. E. Capuj, J. Gomis-Bresco, F. Alzina, A. Pitanti, A. Griol, A. Martínez and C. M. Sotomayor Torres . Scientific Reports \textbf{5}, 15733 (2015). 

\bibitem{daniphononsources}
D. Navarro-Urrios, J. Gomis-Bresco, F. Alzina,  N. E. Capuj, P. D. Garc\'ia, M. F. Colombano, E. Chávez-Angel, and C. M. Sotomayor Torres. J. Opt. \textbf{18}, 094006 (2016). 

\bibitem{downconversion}
M. Hossein-Zadeh and K. J. Vahala, IEEE Photonics Technol. Lett. \textbf{20}(4), 234-236 (2008).

\bibitem{OMOsensing}
W. Yu, W. C. Jiang, Q. Lin, and T. Lu. Nature Communications \textbf{7}, 12311 (2016).

\bibitem{chiweiwong}
X. Luan, Y. Huang, Y. Li, J. F. McMillan, J. Zheng, S.-W.Huang, P.-C. Hsieh, T. Gu, D. Wang, A. Hati, D. A. Howe, G. Wen, M. Yu, G.g Lo, D.-L. Kwong and C. W. Wong. Scientific Reports \textbf{4}, 6842 (2015).

\bibitem{drift}
A. B. Matsko, A. A. Savchenkov, and L. Maleki. Optics Express \textbf{20}, 16234-16244 (2012).

\bibitem{freqstab}
M. Sansa, E. Sage, E. C. Bullard, M. G\'ely, T. Alava, E. Colinet, A. K. Naik, L. G. Villanueva, L. Duraffourg, M. L. Roukes, G. Jourdan and S. Hentz. Nature Nanotechnology \textbf{11}, 552-558 (2016).

\bibitem{thermomechanical}
Chih-Ming Lin, Ting-Ta Yen, V. V. Felmetsger, M. Hopcroft, Jan H. Kuypers, and A. P. Pisano. Appl. Phys. Lett. \textbf{97}, 083501 (2010).

\bibitem{largeamplitude}
B. Yurke, D. S. Greywall, A. N. Pargellis, and P. A. Busch. Phys. Rev. A \textbf{51}, 4211 (1995).


\bibitem{laserlock1}
V. Annovazzi-Lodi, S. Donati, and M. Manna, IEEE J. Quantum Electron. \textbf{30}, 1537 (1994).

\bibitem{laserlock2}
Y. Liu, P. Davis, Y. Takiguchi, T. Aida, S. Saito, and J.-M. Liu, IEEE J. Quantum Electron. \textbf{39}, 269 (2003).

\bibitem{micromaser}
Y.-Y. Liu, J. Stehlik, M. J. Gullans, J. M. Taylor, and J. R. Petta. Phys. Rev. A \textbf{92}, 053802 (2015).

\bibitem{electroniclock1}
B. Razavi. IEEE J. Solid-State Circuits \textbf{39}, 1415 (2004).

\bibitem{electroniclock2}
A. Mirzaei, M. E. Heidari, R. Bagheri, S. Chehrazi and A. A. Abidi, IEEE J. Solid-State Circuits \textbf{42}, 1916 (2007).

\bibitem{organpipes}
M. Abel, K. Ahnert and S. Bergweiler. Phys. Rev. Lett. \textbf{103}, 114301.

\bibitem{weig}
M. J. Seitner, M. Abdi, A. Ridolfo, M. J. Hartmann, and E. M. Weig. Phys. Rev. Lett. \textbf{118}, 254301 (2017).

\bibitem{micromechanical}
D. Pu , R. Huan, and X.Wei. AIP Advances \textbf{7}, 035204 (2017).

\bibitem{superconducting}
D. Markovi\'{c}, J. D. Pillet, E. Flurin, N. Roch, and B. Huard. Phys. Rev. Applied \text{12}, 024034 (2019).

\bibitem{circadian}
J. F. Duffy and C. A. Czeisler. Sleep Med. Clin. \textbf{4}, 165-177 (2009).

\bibitem{adler}
R. Adler. Proc. IRE \textbf{34}, 351-357 (1946).

\bibitem{opticaldrive1}
M. Hossein-Zadeh and K. J. Vahala. Appl. Phys. Lett. \textbf{93}, 191115 (2008).

\bibitem{opticaldrive2}
S. Y. Shah, M. Zhang, R. Rand, and M. Lipson. Phys. Rev. Lett. \textbf{114}, 113602 (2015).

\bibitem{opticaldrive3}
K. Shlomi, D. Yuvaraj, I. Baskin, O. Suchoi, R. Winik, and E. Buks. Phys. Rev. E \textbf{91}, 032910 (2015).

\bibitem{pitanti}
A. Pitanti, J. M. Fink, A. H. Safavi-Naeini, J. T. Hill, C. U. Lei, A. Tredicucci and O. Painter. Optics Express \textbf{23}, 3196 (2015).

\bibitem{optica}
C. Bekker, R. Kalra, C. Baker, and W. P. Bowen. Optica \textbf{4} 1196-1204 (2017).

\bibitem{ILacoustic}
K. Huang and M. Hossein-Zadeh. Optics Express \textbf{26}, 8275 (2018).

\bibitem{SP}
T. J. Johnson, M. Borselli and Painter. Opt. Express, \textbf{14}, 817 (2006).

\bibitem{synchro}
M. F. Colombano, G. Arregui, N. E. Capuj, A. Pitanti, J. Maire, A. Griol, B. Garrido, A. Martinez, C. M. Sotomayor-Torres and D. Navarro-Urrios. Phys. Rev. Lett. \textbf{123}, 017402 (2019).

\bibitem{hopf}
T. Van Vaerenbergh, M. Fiers, J. Dambre, and P. Bienstman. Phys. Rev. A \textbf{86}, 063808 (2012).

\bibitem{naturecomm}
D. Navarro-Urrios, N. E. Capuj, M. F. Colombano, P. D. Garc\'ia, M. Sledzinska, F. Alzina, A. Griol, A. Mart\'nez and C. M. Sotomayor-Torres . Nature Communications \textbf{8}, 14965 (2017).

\bibitem{modetuning}
D. Navarro-Urrios, J. Gomis-Bresco, N. E. Capuj, F. Alzina, A. Griol, D. Puerto, A. Mart\'inez and C. M. Sotomayor-Torres. Journal of Applied Physics \textbf{116}, 093506 (2014).

\bibitem{arnold}
V. I. Arnold. Geometrical Methods in the Theory of Ordinary Differential Equations. (Springer, 2nd Edition,1988)

\bibitem{pikovsky}
A. Pikovsky, M. Rosenblum and J. Kurths, Synchronization: A Universal Concept in Nonlinear Sciences (Cambridge University Press, Cambridge, UK, 2003).

\bibitem{aplwong}
J. Zheng, Y. Li, N. Goldberg, M. McDonald, X. Luan, A. Hati, M. Lu, S. Strauf5, T. Zelevinsky, D. A. Howe and C. W. Wong. Appl. Phys. Lett. \textbf{102}, 141117 (2013).

\bibitem{switching}
J. Maire,  G. Arregui, N. E. Capuj, M. F. Colombano, A. Griol, A. Mart\'inez, C. M. Sotomayor-Torres and  D. Navarro-Urrios. APL Photonics \textbf{3}, 126102 (2018).

\bibitem{timekeeping}
X. S. Yao and L. Maleki.J. Opt. Soc. Am. B \textbf{13}, 1725 (1996).

\bibitem{neuromorphic}
Hoppensteadt, F. C. and Izhikevich, E. M. IEEE Trans. Circuits Syst. I, Reg. Papers \textbf{48}, 133 (2001).

\scriptsize


\end{thebibliography}

\begin{thebibliography}{11}

\bibitem{CPT}
A. W. Snyder and J. D. Love, \emph{Optical waveguide Theory}, Chapman and Hall, (1983).

\bibitem{Vahala}
K. J. Vahala. \emph{Optical microcavities}. Nature (London) \textbf{424}, 839 (2003).

\bibitem{siliconsecondorder}
M. Cazzanelli and J. Schilling. \emph{Second order optical nonlinearity in silicon by symmetry breaking}. App. Phys. Rev. \textbf{3}, 011104 (2016).

\bibitem{nonlinear1}
T. J. Johnson, M. Borselli and O. Painter. \emph{Self-induced optical modulation of the transmission through a high-Q silicon microdisk resonator}. Opt. Express, \textbf{14}, 817 (2006).

\bibitem{nonlinear2}
P. E. Barclay, K. Srinivasan and O. Painter. \emph{Nonlinear response of silicon photonic crystal microresonators excited via an integrated waveguide and fiber taper}. Opt. Express, \textbf{13}, 801 (2005).

\bibitem{nonlinear3}
N. Cazier, X, Checoury, L.-D. Haret and P. Boucaud. \emph{High-frequency self-induced oscillations in a silicon nanocavity}. Opt. Express, \textbf{21}, 13626 (2013).

\bibitem{betaTPA}
M. Dinu, F. Quochi and H. Garcia, \emph{Third-order nonlinearities in silicon at telecomwavelengths}. Appl. Phys Lett. \textbf{82}, 2954 (2003).

\bibitem{sigma}
 R. A. Soref and B. R. Bennett. \emph{Electrooptical Effects in Silicon}. IEEE J. Quan. Elec. \textbf{23}, 123 (1987).

\bibitem{alphalin}
T. Tanabe, H. Sumikura, H. Taniyama, A. Shinya and M. Notomi. \emph{All-silicon subgb/s telecom detector with low dark current and high quantum efficiency on chip}.
Appl. Phys. Lett. \textbf{96}, 101103 (2010).

\bibitem{dani_selfpulsing}
D. Navarro-Urrios N. E. Capuj, J. Gomis-Bresco, F. Alzina, A. Pitanti, A. Griol, A. Mart\'{i}nez and C. M. Sotomayor Torres. \emph{A self-stabilized coherent phonon source driven by optical forces}. Scientific reports \textbf{5}, 15733 (2015).

\bibitem{TOeffect}
D. Navarro-Urrios, J. Gomis-Bresco, N. E. Capuj, F. Alzina, A. Griol, D. Puerto, A. Martínez and C. M. Sotomayor-Torres. \emph{Optical and mechanical mode tuning in an optomechanical crystal with light-induced thermal effects}. Journal of Applied Physics \textbf{116}, 093506 (2014).

\bibitem{cavityoptomechanics}
M. Aspelmeyer, T. J. Kippenberg and F. Marquardt. \emph{Cavity optomechanics}. Rev. Mod. Phys. \textbf{86}, 1391 (2014).

\bibitem{pikovsky}
A. Pikovsky, M. Rosenblum and J. Kurths, Synchronization: A Universal Concept in Nonlinear Sciences (Cambridge University Press, Cambridge, UK, 2003).

\end{thebibliography}
\end{document}